\documentclass[conference,compsoc]{IEEEtran}
\IEEEoverridecommandlockouts

\usepackage{verbatim}
\usepackage{multirow}
\usepackage{caption}
\usepackage{subcaption}
\usepackage{enumitem}
\usepackage{lscape}
\usepackage{rotating}
\usepackage{tabularx}
\usepackage{hyperref}
\usepackage[dvipsnames]{xcolor}

\newcommand{\ada}[1]{\textsf{\textbf{\textcolor{purple}{[[ARL: #1]]}}}}

\newcounter{rcounter}
\newenvironment{rec}[1][]{\refstepcounter{rcounter}\par\itshape\medskip
   \noindent\textit{\textbf{Recommendation \thercounter:} #1}}{\medskip}

\AtBeginDocument{%
  }

\begin{document}

\title{SoK: Technical Implementation and Human Impact of Internet Privacy Regulations}

%\author{\IEEEauthorblockN{Author(s) anonymized for review}}
%\IEEEauthorblockA{}%\textit{Computer Science Department} \\
%\textit{Pomona College}\\
%eleanor.birrell@pomona.edu}

\makeatletter
\newcommand{\linebreakand}{%
\end{@IEEEauthorhalign}
\hfill\mbox{}\par
\mbox{}\hfill\begin{@IEEEauthorhalign}
}
\makeatother

%\begin{comment}
\author{
\IEEEauthorblockN{Eleanor Birrell}
\IEEEauthorblockA{\textit{Pomona College}\\
eleanor.birrell@pomona.edu}
\and
\IEEEauthorblockN{Jay Rodolitz}
\IEEEauthorblockA{
	\textit{Northeastern University}\\
	rodolitz.j@northeastern.edu}
\and
\IEEEauthorblockN{Angel Ding}
\IEEEauthorblockA{
\textit{Wellelsey College}\\
%Wellesley, USA \\
zd2@wellesley.edu}
\and
\IEEEauthorblockN{Jenna Lee}
\IEEEauthorblockA{
	\textit{University of Washington}\\
	%Wellesley, USA \\
	jdlee11@uw.edu
	}

\linebreakand
\IEEEauthorblockN{Emily McReynolds}
\IEEEauthorblockA{
	Unaffiliated \\
	ecm7962@nyu.edu
}

\and
\IEEEauthorblockN{Jevan Hutson}
\IEEEauthorblockA{\textit{Hintze Law PLLC} \\
jevan@hintzelaw.com}
\and

\IEEEauthorblockN{Ada Lerner}
\IEEEauthorblockA{
\textit{Northeastern University}\\
%Boston, USA \\
ada@ccs.neu.edu}
}
%\end{comment}


% \author{Angel Ding}
% \affiliation{%
%   \institution{Wellesley College}
%   %\streetaddress{1 Th{\o}rv{\"a}ld Circle}
%   \city{Wellesley}
%   \state{MA}
% \email{zd2@wellesley.edu}

% \author{Jevan Hutson}
% \affiliation{%
%  \institution{Hintze Law PLLC}
%  \city{Seattle}
%  \state{WA}
%  \country{USA}
% }
% \email{jevan@hintzelaw.com}

% \author{Emily McReynolds}
% %\authornote{Both authors contributed equally to this research.}
% \email{email@email.com}
% %\orcid{1234-5678-9012}
% %\author{G.K.M. Tobin}
% %\authornotemark[1]
% %\email{webmaster@marysville-ohio.com}
% \affiliation{%
%   \institution{Mystery Inc}
%   %\streetaddress{P.O. Box 1212}
%   %\city{Dublin}
%   %\state{Ohio}
%   %\postcode{43017-6221}
% }

% \author{Ada Lerner}
% \affiliation{%
%   \institution{Northeastern University}
%   \city{Boston}
%   \state{MA}
% \email{ada@ccs.neu.edu}

% \author{Eleanor Birrell}
% \affiliation{%
%  \institution{Pomona College}
%  %\streetaddress{Rono-Hills}
%  \city{Claremont}
%  \state{CA}
%  \country{USA}
% }
% \email{eleanor.birrell@pomona.edu}

\maketitle

\thispagestyle{plain}
\pagestyle{plain}


\begin{abstract}
Growing recognition of the potential for exploitation of personal data and of the shortcomings of prior privacy regimes has led to the passage of a multitude of new online privacy regulations. Some of these laws---notably the European Union’s General Data Protection Regulation (GDPR) and the California Consumer Privacy Act (CCPA)---have been the focus of large bodies of research by the computer science community, while others have received less attention. In this work, we analyze a set of Internet privacy and data protection regulations drawn from around the world---both those that have frequently been studied by computer scientists and those that have not---and develop a taxonomy of rights granted and obligations imposed by these laws. We then leverage this taxonomy to systematize 270 technical research papers published in computer science venues that investigate the impact of these laws and explore how technical solutions can complement legal protections. Finally, we analyze the results in this space through an interdisciplinary lens and make recommendations for future work at the intersection of computer science and legal privacy. 
\end{abstract}


\section{Introduction}

Privacy law is shifting and developing rapidly throughout the world. As of March 2022, 157 nations have enacted privacy laws, with 12 laws enacted in 2021 alone~\cite{greenleaf2022now}. Over the past decade,  a large body of computer science research has emerged which studies these laws and their effects. This work has covered topics including rates of corporate compliance; people's ability to exercise privacy rights; design patterns that subvert the consent process; and citizens' expectations, understandings, and actions in the face of laws like the GDPR, which has been called ``one of the strictest privacy laws in the world''~\cite{Solove2017Beyond}.

What have computer scientists learned about these laws and their effects on privacy? And perhaps more importantly, what directions for future computer science research will be most impactful and effective at informing and driving better privacy laws that create more meaningful and equitable privacy outcomes? This paper sets out to answer these questions through an interdisciplinary lens by an authorship team consisting of both computer scientists and legal scholars.

We begin with a close reading of 25 Internet privacy and data protection regulations, which provides a broad overview of a landscape of laws far too numerous to discuss individually yet which often share significant commonalities due to the broad influence of the GDPR on legislators worldwide~\cite{woodward2021countries}. From this close reading we construct a taxonomy of rights guaranteed and business obligations imposed by current Internet privacy regulations.

We then review and systematize the computer science literature around modern digital privacy laws by organizing studies according to the rights and obligations in our taxonomy that they examine. We find that while the research in this space is extensive, its thoroughness varies dramatically, with the overwhelming majority of papers %(X\%) 
studying either California's CCPA or the EU's GDPR, laws which protect only about 6.3\%  of the world's population~\cite{ciaFactbookEU,ciaFactbookWorld,censusCalifornia}. Additionally, certain aspects of these laws such as design patterns in consent banners have been explored deeply, while other aspects (e.g., the right to non-discrimination) have been studied little or not at all.

Building on these analyses, our discussion presents two major arguments. First, we ask how great a limitation our focus on a few specific laws and a few specific rights is to our broad understanding of privacy law writ large. In other words: can we generalize from our deep study of these few contexts? Based on our systematization of the literature and on our interdisciplinary team's analysis of legal factors that can cause varying privacy outcomes even under similar or identical laws, we argue that we should be wary of generalizing specific results beyond their cultural, temporal, and legal contexts. However, we also find that in combination with scholarship from other fields, this body of work has built a compelling case for certain general ideas about privacy law. Most prominently, we analyze the literature's body of evidence against privacy self-management as a paradigm for privacy regulation, arguing that a strong case exists for the necessity of alternative paradigms if we are to produce effective and equitable privacy regulation in the future. We thus conclude our discussion by presenting a roadmap for computer science research that studies approaches beyond privacy self-management.
%  First, we discuss common misconceptions of these laws and offer patches to update our interpretations, in order to ensure that our studies of these laws and their effects are as accurate and impactful as possible. Second, we discuss work studying privacy law from outside computer science, as cross-pollinating with experts from law, the social sciences, business and economics, and other fields can enhance our work. Finally, we build on our understanding of the state of the field to describe a roadmap for future work in this space, with the goal of ensuring that the insights of computer scientists guide existing and future privacy laws toward definitions, designs, implementations, and rights which create and enhance meaningful privacy across the globe.

The contributions of this paper are:
\begin{enumerate}
    % \item We analyze and describe domestic and international privacy for a computer science audience, providing a detailed reference 
    \item We develop a taxonomy of rights and obligations enacted by modern Internet privacy laws through close readings of global laws.
    \item We systematize the computer science literature in the Internet privacy law space, characterizing its extent, depth, and skew within our taxonomy.
    %\item We analyze the generalizability of results in this space through an interdisciplinary lens.% considering the text and context of privacy laws.
    \item We analyze results in this space through an interdisciplinary lens to formulate recommendations for how future computer science research can help guide  more effective and equitable privacy regulation.
\end{enumerate}

\section{Methodology}

This work involved three phases: (1) we identified computer science papers relating to Internet privacy regulations, (2) we developed a taxonomy of rights and obligations imposed by such  regulations around the world, and (3) we systematized the computer science literature relating to these  regulations within this taxonomy, and (4) we formulated recommendations for future work informed by these results. Our methodology for phases (1) and (2) are described in this section. Our systematization  is presented in Section~\ref{sec:legal_systematization}. Our recommendations are discussed in Section~\ref{sec:discussion}.

\subsection{Paper Selection}

%\subsection{Methodology}

We identified ten computer science conferences that regularly published work in the domain of Internet privacy: 
\begin{enumerate}
    \item IEEE Symposium on Security and Privacy (``Oakland'')
    \item ACM Conference on Computer and Communications Security (CCS)
    \item USENIX Security Symposium
    \item Privacy Enhancing Technologies Symposium (PETS)
    \item Symposium on Usable Privacy and Security (SOUPS)
    \item ACM Conference on Human Factors in Computing Systems (CHI)
    \item ACM Conference On Computer-Supported Cooperative Work And Social Computing (CSCW)
    \item Network and Distributed System Security Symposium (NDSS)
    \item ACM The Web Conference (WWW)
    \item ACM Conference on Fairness, Accountability, and Transparency (FAccT)
\end{enumerate}
For each of these ten conferences, one author looked at the title and abstract of each paper published in that venue in the six years between 2017-2022\footnote{For FAccT, which was founded in 2018, we considered the six years 2018-2023.}; we also included papers published in the 2023 conferences through August 2023. This resulted in a preliminary list of 127 computer science papers relating to Internet privacy and data protection regulations.

For each paper on our preliminary list, we applied the same analysis to (1) all of the backwards citations (i.e., works cited by that paper) and (2) all forward citations (i.e., subsequent papers found on Google Scholar that cited that paper). We also iteratively applied this analysis to any additional papers that were published in any of our ten selected venues (e.g., papers that were published prior to 2017 or that were initially excluded). This resulted in a maximal list of 410 candidate papers.

For each of candidate paper, one of the authors read through the full paper and made a final determination about whether the paper was in scope for this work. Papers were considered in scope if they were computer science papers that evaluated the impact of of an Internet privacy or data protection regulation or if they describe or evaluate a tool to  complement or enhance privacy under such a regulation. Merely mentioning a privacy law in the introduction was insufficient to be considered in scope. We focused exclusively on Internet privacy;  sector-specific privacy laws (e.g., HIPAA) were out of scope. Any papers for which the assigned reader was unsure whether it was in scope were discussed with the full set of authors until a consensus was reached. This resulted in a list of 314 in-scope papers. 

Due to practical limitations, we removed any papers published outside of major computer science conferences unless they mentioned a relevant regulation or a specific right under such a regulation in the title of the paper. This resulted in a revised full list of 270 papers: 134 papers published in one of our ten selected venues, 19 papers published in other major computer science conferences (e.g., IMC, VLDB), 102 papers published in other computer science venues (e.g., minor conferences, workshops, or journals), 5 whitepapers, and 11 pre-prints posted on arXiv. 

\begin{table*}
\begin{tabular}{l|l|l|l|l|l}
%\multicolumn{2}{c}{Privacy Self-Management Rights} & \multicolumn{1}{c}{\multirow{2}{4cm}{\centering Absolute Privacy Rights}} & \multirow{2}{4cm}{\centering Business Obligations} \\
%\multicolumn{1}{c}{Elective Privacy Rights} & \multicolumn{1}{c}{Default Privacy Rights} & \multicolumn{1}{c}{} & \\
Definitions & Self-Managed Rights & Fundamental Rights & Business Obligations & Applicability & Enforcement\\
\hline
1. Personal info & 1. Right to access & 1. Right to not be subject  & 1. Notice and Transparency & 1. Subjects& 1. Protection Auth.\\
2. Anonymization & 2. Right to portability  &  to automated decisions  & 2. Purpose/Processing Lim. & 2. Organizations & 2. Gov. Agency\\
& 3. Right to correct & 2. Prohibitions on certain    & 3. Data Minimization & &3. Elected Official\\
& 4. Right to delete  & technologies & 4. Security Requirements & & 4. Priv. Right of Act.\\
& 5. Right to opt-out of  & 3. Prohibitions on certain   & 5. Privacy by Design & & 5. Class Action\\
& processing &processing & 6. Record Keeping & & 6. Arbitration\\
& 6. Right to consent to & 4. Right to nondiscrim.  &  7. Cross-border Trans. Lim. & & 7. Civil Penalties\\
& processing & (based on protected attr.)&  8. Risk Assessment & & 8. Criminal Penalties\\
& & 5. Right to nondiscrim. &  9. Contracting Reqs.&\\
& & (for invoking rights)&  10. Breach Notification&\\
\end{tabular}
\caption{Taxonomy of rights and obligations under Internet privacy and data protection regulations around the world}\label{tab:taxonomy}
\end{table*}

\begin{comment} %% Old table
\begin{table*}
\begin{tabular}{l|l|l|l}
\multicolumn{2}{c}{Privacy Self-Management Rights} & \multicolumn{1}{c}{\multirow{2}{4cm}{\centering Absolute Privacy Rights}} & \multirow{2}{4cm}{\centering Business Obligations} \\
\multicolumn{1}{c}{Elective Privacy Rights} & \multicolumn{1}{c}{Default Privacy Rights} & \multicolumn{1}{c}{} & \\
\hline
1. Right to access & 1. Right to opt-in to cookies & 1. Right to non-discrimination  & 1. Notice and Transparency \\
2. Right to portability  & 2. Right to opt-in for children's data &  2. Right to not be subject to auto-& 2. Legal Basis for Processing \\
3. Right to correct & 3. Right to opt-in for sensitive data & mated decision making & 3. Purpose Limitation \\
4. Right to delete  & & & 4. Data Minimization \\
5. Right to opt-out of processing & & & 5. Security Requirements \\
& & & 6. Privacy by Design \\
& & & 7. Record Keeping \\
& & & 8. Limits on Cross-border Transfers \\
\end{tabular}
\caption{Taxonomy of rights and obligations under Internet privacy and data protection regulations around the world}\label{tab:taxonomy}
\end{table*}
\end{comment}

\subsection{Legal Taxonomy}\label{subsec:methods-taxonomy}

%We analyzed the three laws most commonly referenced laws in the papers on our final list---GDPR, COPPA, and CCPA---along with the updated California regulation scheduled to go into effect in January 2023---California Privacy Rights Act (CPRA)---and four additional regulations from other countries around the world---Canada's Personal Information and Electronic Documents Act (PIPEDA), South Korea's Personal Information Protection Act (PIPA), Singapore's Personal Data Protection Act (PDPA), and Brazil's Lei Geral de Proteção de Dados (LGPD). %These laws include all of the regulations that are currently in effect or that have been enacted that were the subject of papers included in our systematization\footnote{One other paper investigated the impact of PDPB, a data privacy regulation proposed in India. However, that regulation has not been passed into law.}

We organize our central presentation of the literature around a taxonomy of legal features. This taxonomy was derived by the legal scholar members of our team through close readings of 25 laws. These laws comprise all Internet privacy and data protection regulations among countries that represent 75\% of the global population and 75\% of global GPD.
Drawing on the text of these laws, we developed a taxonomy of rights and obligations imposed by current Internet privacy and data protection regulations (Table~\ref{tab:taxonomy}).
\begin{comment}
\begin{enumerate}
    \item Privacy Self-Management Rights
    \begin{enumerate}
        %\item Right to opt-out of collection 
        \item Right to opt-out of processing
        \item Right to access
        \item Right to portability
        \item Right to correct
        \item Right to delete
    \end{enumerate}
    \item Default Privacy Rights
    \begin{enumerate}
        \item Right to opt-in to cookies
        \item Right to opt-in for children's data
        \item Right to opt-in for sensitive data
    \end{enumerate}
    \item Absolute Privacy Rights
    \begin{enumerate}
        \item Right to not be subject to automated decisions
        \item Right to non-discrimination
    \end{enumerate}
    \item Business Obligations
    \begin{enumerate}
        \item Notice/Transparency
        \item Legal Basis for Processing
        \item Purpose Limitation
        \item Data Minimization
        \item Security Requirements
        \item Privacy by Design
    \end{enumerate}
\end{enumerate}
\end{comment}

\subsection{Paper Coding}
We deductively coded each of the 270 papers in our revised full list using our taxonomy of legal features. We also inductively coded each paper for: research methodology, system stage, platform, and applicable laws.

\section{Systematization}\label{sec:legal_systematization}

 % the full set of regulations  representing all Internet privacy regulations from the union of countries representing 75\% of from \todo{XX} jurisdictions. First, we briefly describe all the papers we systematize, presenting those descriptions within a text structure that parallels the taxonomy, describing the legal feature(s) covered by the taxonomic category in the corresponding section. Where a paper covers more than one legal aspect, we described each of those contributions separately, each in the appropriate section. Since the taxonomy describes the legal landscape independent of how it has been viewed or studied by computer science, a significant number of legal features included have been studied in zero CS papers. We nonetheless include sections for these ``empty'' legal features to define the category and emphasize the absence of work on these features as a first-class finding. Second, we describe how the papers fell out on our other taxonomic axes: system stage, methods, platforms, and specific laws studied. \todo{table}? 

Our legal taxonomy organized features of privacy laws into six general categories: (1) definitions, (2) self-managed rights, (3) fundamental rights and prohibitions, (4) obligations, (5) applicability, and (6) enforcement. We also identified  papers that did not study any specific legal features, e.g., by observing changes in tracking before and after a law was implemented but without connecting hypothesized changes in tracking to any particular aspect of the law beyond the fact that it is a privacy law. We describe these papers under a General heading at the end of the section.

\vspace{5pt}
\paragraph{\it \hspace{-35pt} Research Methodologies} 160 papers (59.3\%) conducted measurement studies to observe the implementation of legal requirements. Of those, 60.0\% used automated techniques to conduct large-scale studies and 51.3\% conducted manual measurement studies (some papers performed multiple measurements). 20.6\% of these paper included a longitudinal study that analyzed changes over time, and 16.3\% included a cross-jurisdicational study that analyzed differences between different legal jurisdictions. 

79 papers (29.3\%) used human-computer interaction methods to investigate the human impact of privacy laws. Of these, 51.9\% conducted an exploratory study, 51.9\% conducted a large-scale quantitative study, and 24.1\% conducted an experimental study (some conducted multiple studies). 

45 papers (16.7\%) used systems techniques to implement and evaluate a system for implementing or enhancing privacy regulations. 10 papers (3.7\%) introduced novel attacks based on legal privacy features. 17 papers (6.3\%) used theoretical methods such as cryptography. 15 papers (5.6\%) introduced frameworks. 

\vspace{5pt}
\paragraph{\it \hspace{-34pt} System Stage} Of the 45 papers that about tools and systems, 6 were deployed with a non-trivial userbase in the real world, 23 had implemented prototypes, 2 were in the design phase, and 3 were still high-level proposals. 

\vspace{5pt}
\paragraph{\it \hspace{-35pt} Platform}
34.1\% of papers investigated privacy features specifically in the context of websites, 18.9\% focused on websites, 9.6\% focused on IoT devices, and 13.7\% focused on other domains including databases, routers, TVs, social media, blockchains, networks, and cloud services. 
33.7\% of papers did not apply to any specific platform. 

\vspace{5pt}
\paragraph{\it \hspace{-34pt} Applicable Laws}
87.0\% of papers on our revised final list studied the implementation and impact of GDPR. 17.0\% studies CCPA, 9.3\% studied COPPA, and just 9.6\% studied any other law. Only six papers (2.2\%) considered laws outside of the United States and Europe. Two papers looked at Canada's PIPEDA~\cite{zhang2021whether,narayanan2010myths}. One paper included India's proposed PDPB~\cite{singh2020technical}, one included Singapore's PDPA~\cite{qamar2021detecting}, one paper included the Phillipines's PDPA~\cite{schaffner2022understanding}, and one paper included Turkey's KVKK~\cite{kekulluouglu2023we}. 

%%%%%%%%%%%
%%
%%  Definitions
%%
%%%%%%%%%%%
\subsection{Definitions}

Legal interpretation of privacy regulations depends strongly on interpretations of key definitions. Many works that consider this problem do within the scope of a specific regulatory requirement. However, 8 papers explore definitions of terms that broadly affect the scope of applicable privacy regulations.

Giomi et al.~\cite{giomi2023unified} proposed a statistical framework for GDPR-compliant anonymization. Gomez et al.~\cite{gomez2023sensitive} explored user perceptions of what constituted sensitive information. 
Kutylowski et al~\cite{kutylowski2020gdpr} discuss challenges induced by GDPR's definitions of personal data and pseudoanonymization when applying GDPR's requirements in real-world systems.  Cohen and Nissim (2020)~\cite{cohen2020towards} formalize GPDR's ``singling out'' terminology---used to define identifiable data in Recital 26---as a privacy attack they term ``predicate singling out'', then establish the relationship between this definition and existing techniques such as differential privacy and $k$-anonymity.

Cohen~\cite{cohen2022attacks} demonstrated a set of powerful attacks that show that common anonymization techniques such as $k$-anonymity might fall short of GDPR's legal standard for de-identifying data. Narayanan and Shmatikov~\cite{narayanan2010myths} also discuss the challenges of applying algorithmic technologies to legal definitions of personally-identifiable and de-identified information.

% GDPR anonymization
Gruschka et al.~\cite{gruschka2018privacy} observe that GDPR's rules about the storage and processing of personal information do not apply to anonymous data, but that anonymizing data sets is a difficult problem; they analyze two projects that analyze sensitive large datasets to identify how those projects attempt to protect users' privacy as required under GDPR. %They identify five techniques employed by these projects---explicit consent, security of the processing system, (pseudo)anonymization, processing biometric templates, and limited storage duration---but observe that significant differences exist between the strategies adopted by the two projects. 
% post-GDPR recommendations with pre-GDPR data

%Lucaj et al~\cite{lucaj2023ai} consider the impact assessment requirement the proposed EU AI Law would impose on ``high-risk'' AI systems. The find that a regulatory efforts would be more effective if they focus on practices throughout the AI lifecycle to determine impact, rather that focusing exclusively on outcomes. %scope

%Amaral et al.~\cite{amaral2023nlp} use NLP techiques to automatically determine whether data processing agreements are consistent with GDPR.

%%%%%%%%%%%
%%
%%  Self-managed rights
%%
%%%%%%%%%%%
\subsection{Self-Managed Rights}

Many Internet privacy and data protection regulations introduce rights that individuals can invoke to protect their personal information. These include rights to access, rights to portability, rights to correct, rights to delete, and rights to opt-out of certain types of processing. 

\subsubsection{Right to Access}\label{sec:analysis:rightToAccess}
%Maps directly to Section~\ref{sec:analysis:rightToAccess}.

11 laws we examined provide people with a right to access, although the detailed requirements (e.g., timing requirements, format specifications, etc.) and exceptions to the right of access vary between laws. 
This right has a high level of awareness in the EU~\cite{kuebler2021right,petelka2022generating}; people are less aware of access requests in other jurisdictions, even those with legally-guaranteed rights of access, although many companies honor access requests regardless of user location~\cite{petelka2022generating}. 

%%%%%%%%%%%%%%%%%% SAR mechanisms %%%%%%%%%%%%%%%
\vspace{5pt}
\paragraph{\it \hspace{-30pt}Access Request Mechanisms} 
While researchers have consistently found that the most common Subject Access Request (SAR) mechanisms are email requests and web form~\cite{urban2018unwanted,urban2019study,bufalieri2020gdpr}, detailed procedures are generally different for every website~\cite{bufalieri2020gdpr}. Moreover, many websites~\cite{bufalieri2020gdpr} and child care apps~\cite{gruber2022we} do not accept requests via email, although user studies suggest people consider email the most natural way to claim their right to access data~\cite{alizadeh2020gdpr}.

%%%%%%%%%%%%%%%%%% SAR authentication %%%%%%%%%%%%%%%%%%
\vspace{5pt}
\paragraph{\it \hspace{-30pt}Authentication of Access Requests}
Since responses to SARs can contain personal information, returning data without authenticating the request can constitute a data breach. However, refusing a genuine request denies people their right to access. This tension makes authentication critical for privacy of SARs.  Several research projects have attempted to quantify how organizations authenticate SARs~\cite{boniface2019security,cagnazzo2019gdpirated,di2019personal,bufalieri2020gdpr,di2022revisiting} with somewhat varying results: an estimated 10-71\% use national ID cards to authenticate requests, 15-31\% use subject email access, 15-36\% use subject account login, 6-22\% use secret questions or confidential information, 0-11\% use device cookies, and 1-5\% call the data subject. Researchers have also examined whether companies consider IP addresses as sufficiently identifying to authenticate a request~\cite{adhatarao2021ip}, with negative results. Boniface et al. examined recommendations from the 28 EU countries' Data Protection Authorities (DPAs) and found inconsistent guidelines~\cite{boniface2019security}. %4 DPAs Authentication recommended requiring a national ID card to authenticate requests. Conversely, two DPAs recommended using any identifying information and three recommended using the least sensitive identifier or following the practice of data minimization.

To evaluate authentication of SARs, several projects have issued experimental, spoofed requests. A 2014 study found 25\% of websites returned information in response to requests from email addresses that didn't match the account information~\cite{herrmann2016obtaining} . Work conducted after GDPR went into effect produced varying results. One study found that 58/334 popular websites failed to take any steps to authenticate requests~\cite{bufalieri2020gdpr}.
Using publicly information, researchers have had success rates of 10/14~\cite{cagnazzo2019gdpirated}, 15/55~\cite{di2019personal}, and 60/150~\cite{pavurgdparrrrr} at accessing PII in response to spoofed SARs.  Social engineering persuaded 2-6\% of companies to accept weaker authentication than initially requested~\cite{di2019personal,bufalieri2020gdpr}. Around half of vulnerable organizations remained vulnerable to the same attack years later~\cite{di2022revisiting}, and  mid-sized organizations and non-profits were most frequently vulnerable~\cite{pavurgdparrrrr}. 

An interview study found that companies don't receive many requests for access (i.e., less than 100 per year at some large companies),  %but receive more for Correction (e.g., 1200 to 1600), 
that some haven't suspected any misuse of SARs, and that some have just 4-5 people with the access required to answer SARs while others handle them across a full customer service department~\cite{di2022revisiting}.

%%%%%%%%%%%%%%%%% SAR Responses %%%%%%%%%%%%%%%%
\vspace{5pt}
\paragraph{\it \hspace{-35pt} Compliance with Right to Access}
A lot of work has examined organizations' compliance with SARs, e.g., by making legitimate requests and analyzing the outcomes of those requests~\cite{herrmann2016obtaining,urban2018unwanted,urban2019study,boniface2019security,bufalieri2020gdpr,kroger2020app,bowyer2022human,samarin2023lessons}.

%%
%% Responses and deadlines
%%

One focus of this work has been quantifying how many organizations violate GDPR's right to access by failing to handle the request within its 30 day deadline. Estimates of this non-compliance rate have varied, with various studies reporting 34-40\% non-compliance~\cite{urban2018unwanted}, 36-46\%~\cite{urban2019study}, 28.2\%~\cite{bufalieri2020gdpr}, and 24.4\%~\cite{bowyer2022human} of websites responding to SARs within the legal deadline. However, some of those responses stated that no data was found. Kr\"oger et al. performed a similar experiment for mobile apps and found that 19-28\% of apps failed to handle the request within 30 days~\cite{kroger2020app}. One study found that a sending a reminder email to the company decreased non-compliance to 20\%~\cite{kroger2020app}. Another study noted inconsistencies in how companies counted the time limit (from initial request or from the time they received additional information required)~\cite{urban2018unwanted}.

%%
%% General Usability 
%%

User studies in which website users~\cite{bowyer2022human,petelka2022generating} or smart home users~\cite{chalhoub2022data} issue requests show people find the process confusing and frustrating and authentication can be difficult.

%%
%% Response format 
%%

Studies have also quantified the distribution of data formats returned in response to SARs. Over half of responses were answered with data in structured  formats such as CSV, JSON, or XML~\cite{bufalieri2020gdpr,kroger2020app,urban2018unwanted}, but many non-machine-readable formats were received as well, including screenshots, pdfs, raw-text emails, and printed files~\cite{bufalieri2020gdpr}. Many of these formats are not what users expect to receive (pdfs, word files, or spreadsheets)~\cite{alizadeh2020gdpr}. Some studies found that data was unintelligible due to obscure labeling or formatting errors~\cite{urban2019study,kroger2020app}. User studies find that people often found the returned data incomprehensible, meaningless, unusable, or not useful~\cite{bowyer2022human,wei2020twitter,petelka2022generating}. Some users also struggled to open unfamiliar file formats (e.g., JSON files)~\cite{veys2021pursuing}.

%%
%% Scope of data returned
%%

Responses to SARs often fall short of user expectations~\cite{alizadeh2020gdpr,bowyer2022human,petelka2022generating}: most users would like responses to include derived data (82\%), data acquired from third-parties (81\%), and metadata (73\%), but only a minority of responses (39\%, 49\%, and 4\% respectively) included these data and the returned data was often incomplete~\cite{bowyer2022human}. Some companies (22\% in 2015 prior to GDPR, 6\% in 2019) returned only data types and not actual values~\cite{kroger2020app}. In 2014, only 43\% of apps and websites returned data that matched the observed data collected~\cite{herrmann2016obtaining}. In 2022, many Android apps do not return data types that can be observed through traffic analysis~\cite{samarin2023lessons}. However, a series of focus groups still found that people are surprised, shocked, and scared by the the level of details that some data downloads provide~\cite{veys2021pursuing,arias2023surprised}, and user reported finding the data Twitter and Facebook shared about tracking to be illuminating~\cite{wei2020twitter}.

%%%%%%%%%%%%%%%%% tools/dashboards %%%%%%%%%%%%%%%%%%%%%
\vspace{5pt}
\paragraph{\it \hspace{-34pt} Dashboards and Tools} 

To address usability shortcomings of current SARs, some companies offer tools or dashboards for directly downloading or interactively exploring data. However, these rarely provide both an online tool and an opportunity to download formatted data, they often omit data of concern to users, and they do not help users understand what data is collected by companies~\cite{urban2019your}. A qualitative examination of 10 privacy dashboards determined that none complied with GDPR's right to access~\cite{tolsdorf2021case}. Nonetheless, a user study focused on Facebook's transparency dashboard found that people were surprised by and changed their attitudes after interacting with the dashboard~\cite{arias2023surprised}. 

Researchers have proposed and developed new tools to support the right to access, including a privacy dashboard intended to facilitate user rights under GDPR~\cite{raschke2017designing},  a co-design effort to propose designs, formats, and tools for future data downloads to achieve transparency goals~\cite{veys2021pursuing}, and a document engineering approach to designing and evaluating a disclosure interface~\cite{norval2022disclosure}. Arfelt et al.~\cite{arfelt2019monitoring} express GDPR's access requirement in temporal logic and find it can be efficiently monitored. Jordan et al. built VICEROY, a browser extension that enables people without accounts to make verifiable consumer requests (CCPA) or subject access requests (GDPR) in order to exercise rights to access, correct, and delete personal information using privacy-preserving techniques~\cite{jordan2021viceroy}. Agarwal et al~\cite{agarwal2021retrofitting} built a tool for supporting access requests in legacy database systems. Luckett et al.~\cite{luckett2021odlaw} built a tool for retrofitting access requests into legacy systems. Shezan et al. built CHKPLUG, a static analysis tool that identifies missing information flows in WordPress Plugins that correspond to incomplete responses to access requests~\cite{shezan2023chkplug}.

\subsubsection{Right to Portability}
%Maps directly to Section~\ref{sec:analysis:rightToPortability}
\label{sec:analysis:rightToPortability}
Eight laws provide a right to portability. Among GDPR's self-management rights, this was the least-known and the most misunderstood~\cite{kuebler2021right}.

% actual requests made
Three independent projects---one conducted in 2018~\cite{wong2018portable} and two conducted in 2021~\cite{syrmoudis2021data,kuebler2021right}---made portability requests to hundreds of online services under GDPR; their results were compatible. First, 25\% to 30\% of companies failed to provide data export within the GDPR-mandated time frame of 30 days. Second, 40\% to 50\% of file formats received were not compatible with GDPR's requirement that data be exported in a structured, commonly-used, and machine readable format~\cite{wong2018portable,wong2019right}. 
Larger companies provide a larger scope of data export, more rigorous procedures for authentication, and are more likely to provide import options~\cite{syrmoudis2021data}, but import options remain generally rare (about 25\% of studied services)~\cite{kuebler2021right}.

Two projects looked at data portability for IoT~\cite{turner2021exercisability,barth2021case}. While most (4/4 and 18/34) devices and platforms returned data within GDPR's 30 day requirement, several issues were encountered including lack of request authentication~\cite{barth2021case}, data not structured for machine readability~\cite{turner2021exercisability,barth2021case}, and lack of documentation and explanation~\cite{turner2021exercisability}. 

% awareness
10\% of participants had considered switching between online services, but about two-thirds felt that lack of portability was an obstacle to switching~\cite{kuebler2021right}. The Odlaw tool~\cite{luckett2021odlaw} also supports portability requests.

\subsubsection{Right to Correct}
%Maps directly to Section~\ref{sec:analysis:rightToCorrectData}
\label{sec:analysis:rightToCorrectData}

Eleven privacy regulations provide a right to correct data.  This right to correct was  one motivation for building VICEROY~\cite{jordan2021viceroy}, an authenticated access tool for privacy self-management rights. However, no work has focused explicitly on the right to correct.

\subsubsection{Right to Delete}
%Maps directly to Section~\ref{sec:analysis:rightToDeletion}. Note this section has subsections which we can't further nest.
\label{sec:analysis:rightToDeletion}
11 laws included some form of a right to delete, including rights to withdraw consent. Similar rights have also been granted by other laws, including the Right to be Forgotten. 
Among GDPR's self-management rights, this right was the most widely known~\cite{kuebler2021right}. 

\vspace{5pt}
\paragraph{\it \hspace{-35pt} The Right to Be Forgotten}
% Delists
Researchers have examined trends among delisting requests, including who makes these requests (most commonly law firms and reputation management services~\cite{bertram2019five} about young adult men~\cite{xue2016right}) and types of information delisted (most commonly personal information or official information relating to illegal activities~\cite{bertram2019five,xue2016right}). While it is possible to use data-driven inferences to find many delisted links, the right to be forgotten does appear to enhance privacy of the delisted material~\cite{xue2016right}. 

\vspace{5pt}
\paragraph{\it \hspace{-30pt}The Right to Erasure}
% User studies
The right to erasure is compatible with users' expectations and normative beliefs about privacy~\cite{farke2021privacy,mohamed2018online,schaffner2022understanding}, but  users often are not aware of or do not use deletion controls~\cite{farke2021privacy,habib2020s}. Many people assume deleted posts are sensitive or compromising, although 80\% of users have deleted a post, often for innocuous reasons~\cite{minaei2022empirical}.

% Usability
Work conducted in 2014 found that 52-57\% of websites and apps in Germany deleted accounts upon request~\cite{herrmann2016obtaining}. 
An analysis of a stratified sample of English-language websites found that 27\% updated or added privacy policies relating to deletion after GDPR came into effect~\cite{habib2019empirical}, although many blockchain systems do not address this right in their policy~\cite{sauglam2020data}. However, users are often not aware of or do not use deletion controls~\cite{farke2021privacy,habib2020s},  sometimes because they cannot locate those controls~\cite{habib2020s,take2022feels} or due because they encountered barriers such as authentication requirements, account requirements, paywalls, and dark patterns~\cite{take2022feels,schaffner2022understanding}. Sometimes controls are insufficiently fine-grained, e.g., only offering an option to delete a full account~\cite{mohan2019analyzing}. People also reported that it was difficult to determine whether removal was successful and that keeping information off the Internet required ongoing efforts~\cite{take2022feels}, and one third of attempted account deletion requests were never complete~\cite{schaffner2022understanding}.

\vspace{5pt}
\paragraph{\it \hspace{-33pt} Challenges to Deletion}

Researchers have identified a number of technical challenges to implementing the right to delete, such as deleting all replicas of data including distributed data centers, backups, and offline copies~\cite{sarkar2018towards,shastri2019seven,shah2019analyzing,de2020can}. %Features of modern systems such as replication (e.g., in Redis subsystems~\cite{shah2019analyzing} or across machines, disks, backups, and data centers in the cloud~\cite{shastri2019seven}). 
High-efficiency lazy deletion algorithms have been found to take hours on database such as Redis~\cite{shah2019analyzing} or months (e.g., 180 days in Google Cloud)~\cite{shastri2019seven} to delete data, and systems are sometimes not built to guarantee that all copies of data were deleted~\cite{shastri2019seven}.
A GDPR-compliant version of Redis which deletes data immediately incurred significant overhead~\cite{shah2019analyzing}, as did a SQL-based implementation designed to automate compliance with deletion and retention requirements~\cite{scope2022harmonizing}. However, some researchers argue that state-of-the-art practices can still implement these rights~\cite{politou2018forgetting,deshpande2021sypse}. Other research suggests that rights to delete might be fundamentally incompatible with some technologies such as blockchains~\cite{sauglam2020data}, although Farshid et al.~\cite{farshid2019design} propose a blockchain-like technology that supports data deletion. 

Legal exceptions can also pose a barrier to the right to delete; some electronic monitoring apps claim exemption from the Right to Delete under CCPA citing clauses about retaining data to ``comply with a legal obligation''~\cite{owens2022electronic}. 

% Tools
\vspace{5pt}
\paragraph{\it \hspace{-35pt} Tools for Deletion}   Garg et al.~\cite{garg2020formalizing} and Godin and Lamontagne~\cite{godin2021deletion} provide two possible formal definitions of deletion. 
Arfelt et al.~\cite{arfelt2019monitoring} express GDPR's right to erasure in temporal logic and find it can be efficiently monitored. 
Agarwal et al.~\cite{agarwal2021retrofitting} built a tool for supporting deletion requests in legacy database systems. Luckett et al.~\cite{luckett2021odlaw} built a tool for adding deletion support to legacy systems. CHKPLUG, a static analysis tool that looks for missing information flows in WordPress Plugins, is designed to identify incomplete responses to deletion requests~\cite{shezan2023chkplug}.

\subsubsection{Rights to Opt-out of Processing}
Seven laws grant a right to opt-out of some processing, with different scope. %although the scope of these rights vary. 

%\vspace{10pt}
%\noindent\textit{The Right to Opt-out of Sale under CCPA}
\vspace{5pt}
\paragraph{\it \hspace{-35pt}  The Right to Opt-out of Sale under CCPA}
While compliance with the CCPA's right to opt-out of sale has increased over time since CCPA went into effect~\cite{o2021clear}, there are still numerous websites that are covered by CCPA but do not provide an opt-out of sale link on their homepage~\cite{o2021clear,van2022setting}. Opt-out interfaces often exhibit dark patterns and other interface designs that make it harder to opt-out~\cite{o2021clear,van2022setting}; dark patterns that frequently occur in the wild---e.g., asymmetric UI, extra clicks, and fillable forms---significantly decrease opt-out of sale rates among California users~\cite{o2021clear}.

% ccpa opt-out of sale
The right to opt-out of sale is widely misunderstood:  only 30.5-61.1\% of Californians correctly identifying which behaviors would be covered by this right~\cite{chen2021fighting}. Websites are also inconsistent about how they interpret ``sale''\cite{chen2021fighting}.

Some efforts have been made to explore ways to improve the usability of the right to opt-out of sale, including standardized icons and taglines~\cite{cranor2020design,habib2021toggles}---a recommendation incorporated into the California Attorney General recommendations and the text of CPRA---and browser extensions that improve visibility of opt-out mechanisms~\cite{siebel2022impact} and set browser headers and privacy signals to automatically opt-out of sale~\cite{zimmeck2020standardizing} (signals that California regulations recognize as valid opt-outs). However, while the purpose of these signals is generally well-understood, rates of website compliance with these signals is currently low~\cite{zimmeck2023usability}. %One line of research has developed a privacy icon and studied taglines to improve visability and awareness of the right to opt-out of sale; some of their recommendations have been adopted by the California Attorney General. Zimmeck and Alicki (2020)~\cite{zimmeck2020standardizing} developed OptMeowt, a Chrome browser extension with the goal of integrating CCPA Do Not Sell opt-out into the browser by automatically setting headers, cookies, and privacy signals; according to a variety of proposed and enacted standards for automatically signaling opt-out per CCPA's regulation stating that such automated opt-outs must be interpreted as valid opt-outs. Siebel and Birrell~\cite{siebel2022impact} developed COA, a chrome browser extension that enhances visibility of the right to opt-out of sale by automatically detecting opt-out links and displacing those links in a privacy banner overlayed on the webpage. 

\vspace{5pt}
\paragraph{\it \hspace{-35pt} Other Rights to Opt-out of Processing}
% tool for finding opt-out links in privacy policies
Nearly 90\% of websites provide opt-out choices for email communications or targeted advertising in their privacy policy, but that these choices are often hard to find and comprehend because of poor readability and the lack of standardized wording~\cite{habib2019empirical}. 
Kumar et al.~\cite{bannihatti2020finding} built a corpus and a model to extract and classify opt-out links in privacy policies. They classify opt-out links as being related to targeted ads, communication, cookies, analytics, sharing with third parties, and ``other''. They observe variations in kinds and frequencies of opt-outs based on the popularity of sites/ranking in Alexa, and find that many privacy policies have no such links. They created a browser extension, Opt-Out Easy, which surfaces these links to users.  Arfelt et al.~\cite{arfelt2019monitoring} express GDPR's withdrawal of consent (Article 7(3)), right to restrict processing (Article 18), and right to object (Article 21) in temporal logic and demonstrate efficient monitoring. Allegue et al.~\cite{allegue2020toward} provide a consent manager for IoT smart homes. Odlaw~\cite{luckett2021odlaw} claims to support the right to object in legacy systems. 

\subsubsection{Rights to Consent to Processing}\label{sec:consent}
%I think this pulls out a bunch of stuff from 3.2.1
Many laws require affirmative consent for certain types of data processing. GDPR requires consent---defined as a freely-given, affirmative act---for any processing of personal information that is not covered by an alternative legal basis, a model that was also adopted by many subsequently laws.   %There has been a significant body of work by computer scientists addressing consent interfaces, much of it focusing on cookies and European law. More generally, several laws require opt-in consent for data collection in the absence of an alternate legal basis, 

\vspace{5pt}
\paragraph{\it \hspace{-34pt} Consent Interfaces} Many efforts have been made to categorize and quantify consent interfaces. Most have focused on cookie banner designs: 

\begin{enumerate}
    \item \textit{Choice Options.} Although GDPR requires free consent, observational studies have consistently found that 20-37\% of banners presented no options or were confirmation-only~\cite{hu2019characterising,degeling2018we,mehrnezhad2022can}. Rates can be higher in certain countries~\cite{degeling2018we}, certain contexts (e.g., porn websites~\cite{vallina2019tales}), or certain modalities (e.g., desktop websites vs. mobile apps~\cite{gunawan2021comparative}). Banners with choices used a variety of mechanisms including binary buttons, sliders, checkboxes, and per-vendor settings; among choice mechanisms, checkboxes were the most common immediately after GDPR went into effect~\cite{degeling2018we}, but options to opt-out of cookies directly in a cookie banner were rare~\cite{degeling2018we,sanchez2019can,utz2019informed,khandelwal2023automated}.

    \item \textit{Location.} Most desktop cookie banners (57.9\%) are implemented as bars at the bottom of the page~\cite{utz2019informed}. 

    \item \textit{Design Elements.} Several studies have found that highlighting, pre-selection, and other forms of nudging and dark patterns are common~\cite{utz2019informed,matte2020cookie,nouwens2020dark,mehrnezhad2022can,reynolds2022analysis}. % and 57.4\% used some form of nudging~\cite{utz2019informed}.  By 2020,  75.9\% of banners with multiple choices highlighted the accept cookies option~\cite{mehrnezhad2022can}. Nouwens et al. (2020)~\cite{nouwens2020dark}  scraped the design of the top 10,000 websites on the five most popular consent management platforms and found out that dark patterns are ubiquitous. Matte et al.~\cite{matte2020cookie} documented cases where websites pre-selected options to opt-in to cookies. 

    \item \textit{Banner Text.} GDPR requires that cookie purposes be disclosed to the user; this requirement is not always met. Studies have found that a non-trivial fraction of banners fail to mention a purpose~\cite{santos2021cookie}, use vague purpose-specification language~\cite{santos2021cookie}, assign incorrect purposes~\cite{bollinger2022automating}, or use biased text with framing~\cite{santos2021cookie}. 
\end{enumerate}
Computer scientists have identified requirements for cookie consent~\cite{santos2019cookie}, characterized dark patterns~\cite{mathur2021makes}, and developed frameworks for evaluating privacy choice interfaces~\cite{habib2022evaluating,feng2021design}.  Overall, estimates suggest that 54-95\% of cookie banners fail to meet the standards imposed by GDPR for free, informed consent~\cite{matte2020cookie,nouwens2020dark,santos2021consent,bollinger2022automating,reynolds2022analysis}, and some researchers argue that there is no natural set of design requirements that satisfy all of GDPR's elements~\cite{gray2021dark}. Notifications that a banner is not compliant with legal requirements do not consistently result in improvements to the banner design~\cite{hennig2022your}.

The passage of legal regulations regarding consent to cookie collection has driven adoption solutions provided by Consent Management Providers (CMPs). CMP adoption is highest among mid-market sites, but is increasingly used by all sorts of sites, with significant jumps in adoption immediately after laws like GDPR and CCPA went into effect~\cite{hils2020measuring}. However, banners produced by CMPs are not necessarily compliant with all relevant jurisdictions~\cite{toth2022dark,stover2022website,reynolds2022analysis}. The default banner produced by many CMPs uses highlighting to nudge user consent~\cite{toth2022dark}, 38\% of CMPs did not support any nudge-free designs~\cite{stover2022website}, and some cookie consent libraries  support cookie banners with no decline option~\cite{degeling2018we}. 

Consent interfaces have also been observed in the context of trackers in mobile apps---42.6\% were confirmation-only~\cite{nguyen2022freely}---smart homes---interfaces are manipulative, frustrating, and lack options~\cite{chalhoub2022data} and withholding or revoking consent is hard~\cite{chalhoub2021did}---Hybrid Broadcast Broadband TVs---some channels lack consent mechanisms or fail to support consent revocation~\cite{tagliaro2023still}---and voice assistants~\cite{seymour2023legal}.

\vspace{5pt}
\paragraph{\it \hspace{-34pt} Impact of Interfaces on Consent Decisions} In 2022, there were 56\% more cookie banners in the EU compared to other jurisdictions~\cite{rasaii2023exploring}, and many European users report banner fatigue~\cite{kulyk2020has}. Nonetheless, interface design---including nudging and dark patterns---can impact consent decisions.%~\cite{utz2019informed,nouwens2020dark,bermejo2021website,habib2022okay,ma2022prospective}.
\begin{enumerate}
    \item \textit{Choice Options.} Removing the opt-out button from the first page has the most effect on consent choices, increasing consent rates by 20-30 percentage points~\cite{nouwens2020dark,habib2022okay}. Users are significantly more likely to consent to cookies when presented with a confirmation-only or binary-choice banner compared to banners with finer-grained options~\cite{utz2019informed}. The exact number of choices  does not have an effect~\cite{machuletz2019multiple}, but the presence of an opt-out on the first page does~\cite{bouma2023us}. 
    \item \textit{Location.} A large-scale study of European users in the wild found that people are three times as likely to interact with banners in the lower-left than with banners at the top and bottom of a page~\cite{utz2019informed}. However, a later study with predominantly-American MTurk users  found that position had no significant effect~\cite{bermejo2021website}.
    \item \textit{Design Elements.} Pre-selection or defaults can significantly increase cookie opt-in~\cite{utz2019informed,machuletz2019multiple} and result in lower recall and more regret~\cite{machuletz2019multiple}. A color-based nudging bar displaying privacy threat is the most effective mechanism at nudging users away from default options~\cite{bermejo2021website}. However, highlighting does not significantly increase opt-in~\cite{utz2019informed,bermejo2021website}. In general, defaults and nudging can  influence the cognitive decision process, particularly for users with lower privacy concerns~\cite{bahirat2021overlooking}. 
    \item \textit{Banner Text.} Many variations in text such as the labels on buttons and whether a banner implies that declining cookies will negatively effect experience have no significant effect on cookie consent~\cite{habib2022okay,bouma2023us}, but loss versus gain framing for the button labels does~\cite{ma2022prospective}. 
\end{enumerate}

\noindent Behavior can also differs significantly between desktop and mobile devices~\cite{utz2019informed,bermejo2021website,rasaii2023exploring}. 
Manipulative consent interfaces are not constrained to cookie consent; some websites use nudging and dark patterns to steer users towards particular behaviors~\cite{mathur2019dark}. Users familiar with computer security are more likely to change defaults~\cite{bermejo2021website}. 

\vspace{5pt}
\paragraph{\it \hspace{-35pt} Tools for Consent:} 
Consent-O-Matic~\cite{nouwens2022consent} and BannerClick~\cite{rasaii2023exploring} automatically answer consent pop-ups based on a user's preferences. CookieEnforcer~\cite{khandelwal2023automated} automatically identifies cookie notice and predicts actions required to disable unnecessary cookies. Chetri et al.~\cite{chhetri2022data} develop a tool for automatically verifying informed consent as a legal basis. van Hofslot et al.~\cite{van2022automatic} explored the feasibility of using NLP to detect non-compliant banner language; they found that BERT and LEGAL-BERT provided 70-97\% accuracy, but that the models were constrained by the small size and distribution of the labeled training data. Hils et al.~\cite{hils2021privacy} conducted a longitudinal study of privacy preference signals.

\vspace{5pt}
\paragraph{\it \hspace{-34pt} Effect of Consent Requirements on Cookies and Tracking} Researchers have applied two distinct methodologies to measure the effect of consent requirements on cookies and tracking: (1) conducting longitudinal studies before and after a law goes into effect, and (2) quantifying non-consensual processing after a law is in effect. 

\textit{1. Longitudinal Studies:} Multiple independent projects have conducted a longitudinal analysis of the effect of GDPR on tracking and third-parties~\cite{degeling2018we,sorensen2019before,hu2019characterising,jannick2019privacy,urban2020measuring,hu2020multi,johnson2021privacy,dambra2022sally}.  % GDPR: general (impact on market concentration)
Although they observed a reduction in usage of third-party web technology and cookies immediately after GDPR went into effect~\cite{johnson2021privacy,hu2019characterising}, there was no significant long-term drop~\cite{johnson2021privacy,degeling2018we}. There were some changes towards increased market concentration in third-party services~\cite{johnson2021privacy,sorensen2019before,jannick2019privacy,hu2020multi,dambra2022sally}, and one study observed  a 40\% reduction in cookie syncing, though the general shape of the ecosystem didn't change~\cite{urban2020measuring}. Another study  observed an uptick in new third-party cookies placed when GDPR went into effect~\cite{hu2020multi}, which might signify adoption of CMPs. 

Rasaii et al.~\cite{rasaii2023exploring} measured the impact of CCPA and LGPD and found that those laws had no effect on cookies.

Research has also looked at tracking by mobile apps. Although there are geodifferences in tracking behavior~\cite{kumar2022large}, there was no significant change in tracking by apps after GDPR went into effect~\cite{kollnig2021before}. It is common for apps to engage in third-party tracking, and very few of those apps obtain consent~\cite{jia2019leaks,zimmeck2019maps,kollnig2021fait,ioannidou2021general}.  %; they investigated some widely-used third-party tracking libraries and found out that 1) trackers are unclear about their use of local storage like cookies or other information saved on user devices, 2) trackers expect app developers to obtain user consent but don't have a robust measure to ensure consent, and 3) trackers provide compliance guidance to GDPR but they are hard to read or comprehend~\cite{kollnig2021fait}. %Jia et al. (2019)~\cite{jia2019leaks} analyze the traffic of 509 top Android apps with an association-mining approach. They find that The result shows that 76.23\% of the apps collect and transmit private data without the subjects consent, and 34.06\% share personally-identifiable information with third-parties. 
Apps continued to share data with tracking companies prior to user consent after Apple introduced  App Tracking Transparency (ATT)~\cite{kollnig2022goodbye}.
% GDPR longitudinal: third-party tracking
%Kollnig et al. (2021)~\cite{kollnig2021before} compared the practices of almost 2 million U.K. apps before and after GDPR went into effect and found no significant changes in tracking practices between 2017 and 2020. 
% GDPR geographic: tracking
%Dambra et al. (2022)~\cite{dambra2022sally} analyzed telemetry data from 250K users located around the world; they documented high rates of tracking, with particular prevalence of the top trackers across many different sites.
% GDPR geographic: third-party tracking/ad libraries
%Kumar et al.~\cite{kumar2022large} documented differences in application features between versions of apps available in different countries; 5.8\% of apps had geodifferences in their use of third-party libraries, with the highest rates in Tunisia, UAE, and Ukraine. 
%Kollnig et al. (2022)~\cite{kollnig2022goodbye} quantified tracking behavior's after the introduction of Apple's App Tracking Transparency (ATT) feature and found that data sharing with tracking companies prior to user consent remained common in 2022. 
%
Ret at al.~\cite{ren2019information} looked at information exposure by IoT devices across jurisdictions; they noted that US devices contact more third-parties compared to UK devices.  

Although most users feel negatively about third-parties, many take no action to prevent tracking~\cite{coopamootoo2022feel}.
Among those who do, most common actions were using  a browser extension or manually deleting cookies; fewer people used a privacy-oriented browser or private browsing mode~\cite{mehrnezhad2022can}.

\textit{2. Processing without consent:}
Websites without banners sometimes set cookies~\cite{carpineto2016automatic,fang2018investigating,hu2019characterising,utz2023comparing}, and 82.5\% of websites that use CMPs set at least one cookie that is not covered by their cookie banner and therefore assume implicit consent~\cite{bollinger2022automating}. Websites with banners set cookies prior to obtaining consent~\cite{matte2020cookie,trevisan20194,nouwens2020dark,bollinger2022automating,sanchez2019can,utz2023comparing,reynolds2022analysis}, a practice that is even more common among EU government websites~\cite{samarasinghe2022tu}. However, there is some evidence that the practice of setting cookies prior to consent decreased after GDPR~\cite{dabrowski2019measuring,libert2018changes,hu2019characterising}. Some websites set cookies even after the user rejects those cookies~\cite{hu2019characterising,bollinger2022automating,matte2020cookie,sanchez2019can}, some use browser fingerprinting to bypass cookie-consent rules~\cite{papadogiannakis2021user}, and 
3.8\% of top websites use cookie respawning~\cite{fouad2022my}.
However, opting out of cookies does result in fewer cookies~\cite{rasaii2023exploring} and can also result in fewer scripts (and corresponding reduced vulnerability to scripting attacks)~\cite{klein2022accept}.

Some of the issues with non-consensual cookies arise from third-party elements that are imported into a site. Many profiling cookies set prior to user consent are actually set by advertising networks~\cite{trevisan20194}  and ghosted cookies~\cite{sanchez2021journey}---cookies set by hierarchically-imported resources---pose a challenge to consent because the website does not have full control over the setting of these first-party cookies.  Only 12.8\% of third-party cookies have a cookie policy that mentions that cookie, and only 5\% include a description of the cookie's purpose in a well-structured table~\cite{fouad2020compliance}. %Trevisan et al. (2019)~\cite{trevisan20194} also observed that many profiling cookies set prior to user consent were actually set by advertising networks and not the website itself.  

Beyond website cookies, 16.7\% of mobile apps share data with third-party trackers prior to user consent and 1.0\% shared data even after a user explicitly declined consent~\cite{nguyen2022freely}. 20.9\% of apps transmit data without displaying a consent dialogue or privacy notice~\cite{koch2022keeping}. 34.4\% of apps share personal information with third-party data controllers without opt-in consent, most commonly sending the AAID~\cite{nguyen2021share}. Leaks of unresettable user identifiers (UUIs) on Android devices can bypassed the permission consent mechanisms~\cite{meng2023post}. Hybrid Broadcast Broadband TV channels track users prior to receiving consent~\cite{tagliaro2023still}. Attacks can extract personal information about other users from an Amazon Echo device~\cite{furey2019can}. Apps practice deceptive uses of legitimate interest to justify data collection without consent in ways inconsistent with user preferences~\cite{kyi2023investigating}. Some websites continue to send market emails without opt-in consent, without an option to revoke consent, or after a user revokes their consent~\cite{kubicek2022checking}. Karami et al. developed a programming language that generates runtime errors if data are used without consent~\cite{karami2022dpl}. Misconceptions about what is required for consent are common, with many developers believing that having a privacy policy supersedes the need to have explicit consent~\cite{nguyen2021share}; many developers also argued that they rely on third-party app builders or SDKs to make their app compliant, and assumed that libraries would implement GDPR-compliant behavior.

\vspace{5pt}
\paragraph{\it \hspace{-34pt} Effect of Consent Requirements for Sensitive Data}
%Maps directly to Section 3.2.3. Right to Opt-in for Sensitive Data.

%But, it looks like that existing section contains more than the 5 papers coded with this category in the spreadsheet, so it looks like we probably have to separate those out into wherever they landed instead.

Nine laws require opt-in consent for  processing  sensitive personal information. Definitions of ``sensitive'' vary between laws and  differ significantly from what users consider sensitive~\cite{gomez2023sensitive}.
 
% pre-GDPR sensitive information
Work conducted prior to GDPR found that Facebook made significant use of sensitive data for targeting ads prior to GDPR and therefore predicted that the regulation should significantly impact data processing by major ad providers~\cite{cabanas2018unveiling}. 
Subsequent work did find significantly lower rates of tracking for sensitive data in Europe compared to other jurisdictions~\cite{dambra2022sally}, however, tracking persisted on sites relating to sensitive personal information including health~\cite{iordanou2018tracing,dambra2022sally}, sexual orientation and preferences~\cite{iordanou2018tracing,vallina2019tales}, religion~\cite{iordanou2018tracing,dambra2022sally}, and politics~\cite{dambra2022sally}. 
74\% of paid versions of apps held the same dangerous permissions as their free versions ~\cite{han2020price}. 

Restrictions on the collection of sensitive personal information can also have unintended negative consequences. Interviews with industry experts working on algorithmic fairness for machine learning revealed that GDPR's restrictions were viewed as prohibitive, and the resulting lack of access to racial data resulted in no longer trying to detect racial bias in their machine learning systems~\cite{andrus2021we}.

\vspace{5pt}
\paragraph{\it \hspace{-34pt} Effect of Consent Requirements for Children's Data}

Nine laws require parental consent for information about children, but the details (e.g., collection versus processing, the age limit for protection, and the requirements for verifying parental consent) vary between regulations. CCPA and CPRA require opt-in consent (resp., parental consent) to sell personal data about children under 16 (resp., under 13). 

Several projects have investigated COPPA compliance by mobile apps. This work has consistently found that many apps violate COPPA in a variety of ways, including using SDKs that prohibit use in child-targeted apps because they collect or share PII~\cite{reyes2018won,ali2020betrayed,feal2020angel,gruber2022we,razaghpanah2018apps}, using libraries without necessary COPPA-compliant parameters~\cite{ali2020betrayed}, not asking for parental consent prior to collection~\cite{gruber2022we,alomar2022developers}, using fingerprinting-alike libraries to bypass parental consent~\cite{ferreira2018investigating}, or using parental consent mechanisms such as age gates and knowledge-based questions that do not satisfy the FTC's requirements for verifiable parental consent~\cite{alomar2022developers}. Behaviors that violate COPPA have also been observed in COPPA-approved and child-oriented websites~\cite{vlajic2018online} and in voice assistant skills in the ``kids'' category~\cite{liao2020measuring}. Overlapping behavior in free and paid versions of apps might also be indicative of practices that violate COPPA~\cite{han2020price}. 
Xie et al.~\cite{xie2022scrutinizing} automatically analyzed use of children's data by IoT devices and found noncompliance in 8/512 skills. 

%Liao et al. (2020)~\cite{liao2020measuring} document 3 voice assistant skills in the ``kids'' category that do not provide a privacy policy, and 137 kids skills that provide general information without providing specifics on what personal data they actually collect; these skills are potentially in violation of COPPA's requirements about parental consent to the collection and use of personal data about children. 

% COPPA developers
Failure to comply with restrictions on processing children's data might be due to barriers to compliance rather than malicious intent. Developers of popular Android children's' apps report issues including a lack of transparency from libraries, a need for data to understand user behavior, and difficulty monetizing apps in age-appropriate ways~\cite{ekambaranathan2021money}. 

Efforts to facilitate COPPA compliance have resulted in tools that leverage dynamic execution and traffic monitoring~\cite{wijesekera2017our} or machine learning~\cite{basu2020copptcha,xie2022scrutinizing} to analyze apps or IoT devices and detect behavior that violates legal regulations. These tools have high accuracy but are not currently in widespread use.

% COPPA user attitudes
In general, COPPA requirements are consistent with parental expectations and social norms, although younger parents are significantly more accepting of data collection~\cite{apthorpe2019evaluating}.  
%Apthorpe et al.~\cite{apthorpe2019evaluating} conducted a user survey to measure how well provisions of COPPA conformed to privacy norms. They surveyed 195 U.S. parents of children between ages 3 and 13 and found that the legal restrictions on data collection imposed by the regulation generally reflected cultural norms regarding data collection: information flows with COPPA-derived transmission principles were significantly more acceptable than other data flows. Data Collection for advertising purposes and collection about children's birthdays were viewed as particularly unacceptable. There were, however, significant differences between sub-populations that might be indicative of future shifts in cultural norms; younger parents and parents who own smart devices were significantly more accepting of data collection. 
% COPPA counterproductive
However, by incentivizing online services to ban users younger than 13, there is also some evidence that COPPA may have reduced privacy for adolescents, who may lie about their age to join platforms, thus aging out of protections for minors before they turn 18~\cite{dey2013profiling}. % minors in some contexts. For example, the economic costs of handling COPPA's requirements about verifiable parental consent have caused most online services---including Facebook---to ban users younger than 13. This results in some young children lying about their age in order to join the platform; which in turn can result in lower protections for older adolescents. The additional public information can be leveraged to construct accurate lists of underage high schools complete with profiles that contain significantly more information than should be available about minors~\cite{dey2013profiling}. %construct at some point later, these profiles show that the user is over 18 (even though the user is in fact a minor) and are included in search results with additional information in their public profiles. By identifying these minors and analyzing their friend lists, the authors were able to construct accurate lists of underage high schools complete with profiles that contain significantly more information than should be available about minors. This attack demonstrates that current child protection laws are insufficient to protect children's online privacy, and that COPPA's requirement of verifiable consent for young children can have a negative impact on children's privacy. 

%%%%%%%%%%%
%%
%%  Fundamental rights
%%
%%%%%%%%%%%
\subsection{Fundamental Rights and Prohibitions}

In contrast with self-managed rights, which must be explicitly invoked by individuals, fundamental rights impose general prohibitions on certain types of behaviors. These rights are relatively rare in current privacy regulations.

%\subsubsection{Right not to be subject to automated decision-making}
%Maps directly to 3.3.2
Eight laws create rights to not be subject to automated decision making. 
%GDPR grants a right to not be subject to automated decisions. % CPRA will introduce such a right. However, only one paper investigated this right. 
Kaushik et al. studied GDPR's version of this right and found it is commonly misunderstood---many people believe users can opt-out in advance---and fails to meet expectations for transparency about automated decisions~\cite{kaushik2021know}. Krishna et al.~\cite{krishna2023towards} formulate the problem of implementing the right to explanation in the context of automated decisions as an optimization problem robust against model updates to accommodate deletion requests. Other work has considered how this right might be implemented for database systems~\cite{shastri13understanding,shastri2019seven}.

Shastri et al.~\cite{shastri13understanding} also discuss how databases can support two additional rights---prohibitions on facial recognition and prohibitions on targeted advertising---that are granted by GDPR. Analogous rights prohibiting certain technologies or processing activities do not appear in other privacy regulations.

%\subsubsection{Discrimination (on the basis of a protected class)}
%empty (section 3.3.1)
GDPR and LGPD grant freedom from discriminatory processing on the basis of sensitive personal information. Four laws, including CPRA, GDPR, LGPD, and China's PIPA grant a right to non-discrimination if a user invokes their rights to privacy self-management. These right has not been the explored in the computer science literature.

%%%%%%%%%%%
%%
%%  Obligations
%%
%%%%%%%%%%%
\subsection{Obligations}

Obligations are procedural requirements that must be satisfied when collecting or processing data. Examples include transparency requirements, legal basis requirements, data minimization requirements, and privacy by design. Some of these legal aspects---specifically transparency and consent as a legal basis---have been the focus of significant bodies of work by computer scientists. Others have not. 
        
\subsubsection{Notice and Transparency}
%Maps to 3.4.1
%Maybe also needs to be discussed?

Transparency was been a key motivation behind many privacy laws. GDPR calls for data to be ``processed lawfully, fairly and in a transparent manner in relation to the data subject'' [5(1)(a)]. CCPA features the ``right to know'' as the first right granted to consumers. Nine other jurisdictions also have transparency requirements. Despite some challenges~\cite{mhaidli2023researchers}, a lot of work has focused on privacy policies as the most common form of notice; some work has also looked at other transparency mechanisms in the context of legal requirements. 

%The regime of privacy law in the United States is one of notice, explaining what data is being collected, and consent. Consent means any freely given, specific, informed, and unambiguous indication of the consumer’s wishes by which the consumer signifies agreement to the processing of personal information relating to the consumer for a narrowly defined particular purpose. While CCPA’s definition may initially look more limited because of its inclusion of a narrowly defined particular purpose because it overlays provisions for opt-out it may in practice be less restrictive than that of GDPR and its counterparts which are similarly defined but in a context of broader rights. [see GDPR Article 7]

%\subsubsection*{Right to Transparent Disclosure} 

% Discussion of the laws that require notice/transparency

%``In European countries, GDPR and the proposed ePrivacy Regulation, with the UK establishing its implementation, stipulate that online service providers are required to inform individuals that tracking technologies are present, what they do and why, and to receive users' consent to use them''~\cite{coopamootoo2022feel}. ``the French data regulator, CNIL levied a 50 million Euro file for a breach of GDPR's transparency requirements, underscoring informed consent requirements concerning data collection for personalized ads''~\cite{reardon201950,CNIL2019}. CCPA also has transparency requirements.

% Summary of papers about the effect of these laws

%%
%% pp changes from GDPR
%%
\vspace{5pt}
\paragraph{\it \hspace{-35pt} Impact of Regulations on Privacy Policies} Several independent longitudinal studies looked at privacy policies before and after GDPR went into effect to understand it's impact on privacy policies~\cite{degeling2018we,habib2019empirical,weir2020needs,linden2020privacy,zaeem2020effect,amos2021privacy,adhikari2023evolution}. An estimated 4.9\% of apps added privacy policies and 50\% updated pre-exising policies~\cite{degeling2018we}---the most widespread changes in the last decade~\cite{amos2021privacy,adhikari2023evolution}---with many adding particular new content (e.g., options regarding deletion~\cite{habib2019empirical}, information about privacy self-management rights~\cite{adhikari2023evolution}, and information about protections for children's data~\cite{zaeem2020effect}) to meet particular GDPR requirements. 38\% of Android developers reported adding or updating the privacy policy for their app~\cite{weir2020needs}. Overall, privacy policy sensitivity increased over time, with spikes corresponding to enactment of privacy regulations~\cite{lovato2023more}. However, not all companies had privacy policies even after GDPR~\cite{degeling2018we,kumar2022large,manandhar2022smart}. Levato et al.~\cite{lovato2023more} studies the impact of a broader range of legislation using a longitudinal corpus from 1997-2018. Cross-jurisdictional analyses have also identified jurisdictional differences between policies in the EU and the US~\cite{kumar2022large,arora2022tale}, with some evidence suggesting that the GDPR reduced data collection~\cite{arora2022tale}. %A study of geodifferences between privacy policies about 2\% had different privacy policies in different countries; most differences were additional clauses to comply with GDPR and CCPA~\cite{kumar2022large}. 

%They found that 4.9\% of websites added privacy policies policies, but that 30\% of websites didn't have a privacy policy even after GDPR went into effect~\cite{degeling2018we}.  Among websites that had pre-existing privacy policies, GDPR prompted the most widespread changes to privacy policies in the last decade~\cite{amos2021privacy}, with 50\% of websites updating  their policies in May 2018 when it was legally required~\cite{degeling2018we}. Although 39.6\% of website policies showed no change in any of 10 privacy measures, there were significant increases in policies that covered protections for children's personal information and data aggregation~\cite{zaeem2020effect}. 27.3\% of websites either updated or added options regarding data deletion after GDPR went into effect, but only 5\% implemented mechanisms to support DNT~\cite{habib2019empirical}. In a user study of 330 Android developers, 38\% reported adding or updating the privacy policy for their app~\cite{weir2020needs}. 

The impact of regulations on the transparency of privacy policies was mixed. Using NLP techniques, researchers found an increase in terms related to GDPR rights~\cite{degeling2018we,linden2020privacy,adhikari2023evolution} and in granularity of disclosures~\cite{arora2022tale}; most apps with different policies in different countries were due to additional clauses relating to GDPR or CCPA~\cite{kumar2022large}. Some longitudinal studies have found that policies became more specific with improved presentation~\cite{linden2020privacy} and simpler and more regularized~\cite{lovato2023more}. However, many privacy policies showed no improvement in any of 10 privacy measures~\cite{zaeem2020effect}. Many studies have found that privacy policies also became longer~\cite{linden2020privacy,amos2021privacy,adhikari2023evolution} and harder to read~\cite{amos2021privacy}. Some policies covered more data use practices at the cost of reduced specificity~\cite{linden2020privacy}, and information about tracking and information sharing with third parties were still frequently missing~\cite{amos2021privacy}. Few developers reported making substantive changes such as adding popup consent dialogues~\cite{weir2020needs}.

%%
%% pp compliance with GDPR
%%
\vspace{5pt}
\paragraph{\it \hspace{-35pt} Compliance with GDPR Transparency Requirements} According to analyses, privacy policies can violate GDPR in five ways: (1) omitting required information~\cite{contissa2018claudette,liepin2019gdpr} (Fan et al.~\cite{fan2020empirical} identify six required notice categories; Vanezi et al.~\cite{yeratziotis2021complicy} identified a list of 89 terms across 7 groups that should be included in privacy policies), (2) describing data use practices that violate legal restrictions~\cite{contissa2018claudette,liepin2019gdpr}, (3) using unclear language~\cite{contissa2018claudette,liepin2019gdpr,oh2021will}, (4) not providing the privacy policy in an accessible location, and (5) inaccurately or incompletely disclosing data practices.   A large body of work has also  been devoted to developing tools that use machine learning and natural language processing techniques for analyzing privacy policies to determine whether they fulfill the transparency requirements imposed by GDPR~\cite{contissa2018claudette,mousavi2018knight,urban2018unwanted,urban2019study,andow2019policylint,andow2020actions,fan2020empirical,fouad2020compliance,razavisousan2021analyzing,yeratziotis2021complicy,oh2021will,liu2021have,elhamdani2021combined,grunewald2021tira,rahat2021automated,qamar2021detecting,ling2022they,ou2022viopolicy}. Many of these tools have been applied to corpuses of post-GDPR privacy policies to determine rates of compliance with GDPR. 

% (1) omitting required information
%\vspace{5pt}
%\begin{enumerate}
%\item 
\textit{1) Omitting Required Information.} Between 8.3\%~\cite{yeratziotis2021complicy} and 23.7\%~\cite{fan2020empirical} of website privacy policies are missing at least one required category.
Observed compliance levels significantly for different disclosure requirements imposed by GDPR~\cite{rahat2021automated}: more than 90\% of policies disclose categories of data collected and purposes of data processing (albeit sometimes bundled~\cite{mohan2019analyzing}), but only 15.3\% disclose how personal information is used for automated decision making or profiling~\cite{turner2021exercisability}. 43\% of child care apps do not mention processing sensitive data about children~\cite{gruber2022we}. 56\% of privacy policies for browser extensions omit one or more pieces of information required by GDPR~\cite{ling2022they}. Users struggle to identify policy excerpts relevant to GDPR's articles~\cite{mousavi2018knight}.

%Researchers have also looked at consent to privacy policies~\cite{oh2021will,kariryaa2021understanding}---12.3\% of websites provided a button for the user to explicitly opt-in to the terms of the privacy policy and only 10.1\% of sites had separate opt-ins for the privacy policy and the terms of service;  60\% of browser extension users never read their privacy policies, rendering users unable to make an informed choice about browser extensions.

% (2) data use practices that violate legal restrictions
%\vspace{6pt}
\textit{2) Illegal Data Use Practices} None of the papers we included measured prevalence of illegal practices. 

% (3) unclear language
%\vspace{6pt}
\textit{3) Unclear Language.} An estimated 1.4\% of website privacy policies have readable privacy policies~\cite{oh2021will}. Many use vague language~\cite{mohan2019analyzing,ioannidou2021general}. Moreover,  7.6-18.1\% of privacy policies for Android apps contain contradictions that may be indicative of misleading statements~\cite{andow2019policylint,andow2020actions,bui2021consistency}; many are contradictions due to inconsistencies with GDPR's definition of personal information. GDPR requires that privacy notices be understandable to children if a company processes children's data; however, misconceptions about data collection, storage, and processing remain pervasive among children~\cite{sun2021they}, suggesting that current implementations fall short of meeting this legal standard. 

% (4) inaccessible location
%\vspace{6pt}
\textit{4) Inaccessible Policy.} Most websites provide a privacy policy in an accessible location~\cite{oh2021will}, but there still are some without privacy policies~\cite{utz2023comparing}.  Compliance  can be lower in other contexts:   74\% of IoT producers' websites~\cite{turner2021exercisability,manandhar2022smart}, 62.2\% of Google Assistant actions~\cite{liao2020measuring},  50.5-55\% of Android App Store pages~\cite{zimmeck2019maps,han2020price},  32\% of browser extensions~\cite{ling2022they}, 9\% of Smart Home Devices~\cite{manandhar2022smart}, 27.7\% of Amazon Alexa skills~\cite{liao2020measuring}, and only 16\% of porn sites~\cite{vallina2019tales}. Moreover, there aren't clear standards for accessible means for apps and smart home devices~\cite{zimmeck2019maps,manandhar2022smart}.

% (5) inaccurate policy
%\vspace{6pt}
\textit{5) Inaccurate Policy.} App behavior is not always consistent with notices provided by websites and apps ~\cite{reardon201950,andow2020actions,jia2019leaks,zimmeckEtAlCompliance2017,fan2020empirical,bui2021consistency,alomar2022developers}:  42.4-77.9\% of apps exhibit at least one behavior inconsistent with their privacy policy~\cite{andow2020actions,fan2020empirical,xiao2023lalaine} and 17-18\% share information with third parties without disclosing it~\cite{zimmeckEtAlCompliance2017,feal2020angel}, and  app behavior is frequently inconsistent with app privacy  labels~\cite{koch2022keeping,xiao2023lalaine}. 35\% of European websites collect data not disclosed by their privacy policy~\cite{ou2022viopolicy}. 11/165 popular trackers with opt-out choices exhibited data practices inconsistent with their privacy policy~\cite{bui2022opt}. IoT skills for Amazon and Google devices have also been found to be inconsistent with their privacy policies~\cite{guo2020skillexplorer,xie2022scrutinizing}, and 31\% of IoT companion apps share data without disclosing it. Independent work developing a testbed for IoT devices and found that half of 11 sample devices collected data not disclosed by their privacy policy~\cite{subahi2018ensuring}.  Ling et al.~\cite{ling2022they} identified inconsistencies between privacy policies and actual data practices for almost half of browser extensions. 
%\end{enumerate}

%\vspace{5pt}
% overall
Overall, early estimates suggested up to a third of the privacy policies for large companies are not compliant with GDPR's requirements~\cite{contissa2018claudette}, but later work found that number could be as high as 97\%~\cite{rahat2021automated}, with many privacy policies having multiple compliance issues~\cite{liu2021have}. Compliance is higher for top tracking companies, most of which meet the minimum legal requirements set out by GDPR~\cite{urban2019study}. %later work---which considered broader classes of companies and which drew insights from subsequent guidelines and legal rulings---again found rates of compliance with GDPR as low as 3\%~\cite{rahat2021automated,liu2021have}. %Liu et al.~\cite{liu2021have} identified 1,180 compliance issues across 304 privacy policy documents according to their analysis, and Rahat et al.~\cite{rahat2021automated} found that only 3\% of privacy policies from top websites fully comply with GDPR.  
%GDPR also requires that privacy notices be understandable to children if a company processes children's data. However, misconceptions about data collection, data storage, and data processing remain pervasive among children~\cite{sun2021they}, suggesting that current implementations---which can be text buried in existing privacy policies---fall short of meeting this legal standard. 

%%
%% pp compliance with other laws
%%
\vspace{5pt}
\paragraph{\it \hspace{-34pt} Compliance with Other Transparency Requirements} Three projects focused on the transparency requirements imposed by COPPA on apps that target children under 13. Early work found that only 10.8\% of these apps targeted at young children provided a privacy policy in their Google Play Store page in 2013 even though half collected some amount of personal information~\cite{liccardi2014can}. Five years later, only half of parental control apps clearly informed users of their data collection and sharing practices, and only 24\% provided a complete list of third-parties with which they share information~\cite{feal2020angel}. Some companies violate COPPA by failing to address children's data in their privacy policy~\cite{razaghpanah2018apps} or not disclosing data practices in the privacy policy~\cite{zimmeck2019maps}. In 2022,  many child care apps still failed to disclose use of trackers or processing of sensitive data about children~\cite{gruber2022we}. Dempsey et al~\cite{dempsey2018designing} surveyed children ages 7-11 to understand whether it is possible to meet GDPR's requirement to meet GDPR's child-related transparency requirements; the found clear saftey concerns, but also found that children weren't aware that personal data had value. 

% CCPA transparency
Chen et al. found that while almost all U.S. websites describe their data sharing practices, only 24.4\% of those list every category of personal information that is shared with each category of recipient as required by CCPA~\cite{chen2021fighting}. %Most policies provide examples of information categories or recipients or provide separate lists of information categories and recipients without specifying the association; 20.7\% of those policies refer generically to ``personal information'' when describing sharing practices without including a list of personal information that is shared. 
%They also found that policies are inconsistent about how they interpret CCPA's definition of a ``sale''. In a user study with 364 Californian consumers, they found that users do not understand how terms such as ``sale'', ``business purpose'', and ``service provider'' are defined by CCPA and that their understanding of data use practices was impeded by inconsistent usage of terminology across different privacy policies. %They recommend further clarification in future guidelines along with strengthened enforcement to enhance transparency about data sharing. 
Musa et al.~\cite{musa2022atom} proposed a technique for automatically inferring data sharing relationships and validated it against CCPA's data broker registry, suggesting a means to verify accuracy of CCPA disclosures.

% GDPR and PDPA
Qamar et al.~\cite{qamar2021detecting} developed a tool that measures similarity between the text of a law and  a privacy policy; their tool supports compliance evaluation for Singapore's PDPA, but they have not published compliance rates.

%Qamar et al.~\cite{qamar2021detecting} studied privacy policies under the GDPR and the PDPA (Singapore) with the goal of automatically detecting compliance of privacy policies with these laws. They took the approach of having law students manually label policies using the annotation scheme of Wilson et al. (\cite{wilson2016creation}) which annotates data practices as belonging to 10 categories (e.g., first party collection/use, third party collection use, etc.), then trained logistic regression and SVM models, as well as existing language models, to classify privacy policy text. They also applied LDA to group similar segments of the GDPR in preparation for labeling, and manually labeled the PDPA directly because it is shorter. They measure compliance by measuring the similarity between the laws' texts and the privacy policy texts, thresholded based on a small manually labeled set. Primarily the evaluate the properties of their NLP algorithms and models rather than actual compliance with the
%%
%% End: pp compliance with other laws
%%

%%
%% tools to improve transparency
%%
\vspace{5pt}
\paragraph{\it \hspace{-34pt}  Tools to Improve Transparency} One approach to facilitating compliance with transparency requirements is to automatically generate policies that comply with the various laws. Such tools have been developed for COPPA~\cite{liccardi2014can,zimmeck2021privacyflash}, GDPR~\cite{gerl2018lpl,amariles2020compliance,zimmeck2021privacyflash,basin2018purpose,shezan2022nl2gdpr}, CCPA~\cite{zimmeck2021privacyflash}, and CalOPPA~\cite{zimmeck2021privacyflash}. There are also questionnaire-based generators that create privacy policies for mobile apps~\cite{APPGenerator,iubenda,termly,TermsFeed}; three of these claim to generate policies that are compliant with COPPA, GDPR, and CCPA. 
%To facilitate COPPA-compliance, Liccardi et al.~\cite{liccardi2014can} developed a tool for automatically generating privacy policies based on the developer's specification of personal data used by the app.
%Gerl and Meier~\cite{gerl2019layered} used formal language techniques to create a LPL Policy Creator to help businesses create GDPR-compliant privacy policies, and a LPL Policy Viewer to present policies to users. 
%Amariles et al.~\cite{amariles2020compliance} reviewed literature and produced a ``road map'' of future steps around automatically interpreting and generating GDPR-compliant privacy documents. Today, there are four popular questionnaire-based generators that create privacy policies for mobile apps: App Privacy Policy Generator~\cite{APPGenerator}, iubenda~\cite{iubenda}, Termly~\cite{termly}, and TermsFeed~\cite{TermsFeed}. Three of these generators---iubenda, Termly, and TermsFeed---claim to generate policies that are compliant with GDPR, COPPA, and CCPA. 
However, an independent analysis~\cite{zimmeck2021privacyflash} found that the resulting policies are only actually compliant with GDPR as of January 2021; all generated policies violated multiple transparency-related requirements imposed by COPPA and by CCPA. %generated policies were missing a statement about whether children were able to make personally-identifiable information public (COPPA), were missing contact information for all operators that handle personally-identifiable information about children (COPPA), did not handle special requirements for businesses that handle personally-identifiable information for more that 10,000,000 per year (CCPA), and did not meet industry accessibility standards (CCPA). iubenda-generated policies also failed other transparency requirements under CCPA, and Termly-genrated policies also failed additiona requirements under both COPPA and CCPA. 

An alternative approach is to enhance transparency with summarization or annotation. Several independent projects applied NLP to automatically identify and highlight parts of privacy policies are relevant to GDPR's requirements~\cite{tesfay2018read,tesfay2018privacyguide,mousavi2018knight,chang2019automated,arora2022tale}. Mustapha et al.~\cite{mustapha2020privacy} provide an improved tool for automatically annotating privacy policies. %Tesfay et al. (2018)~\cite{tesfay2018read,tesfay2018privacyguide} used machine learning to summarize privacy policies using a risk-based approach to draw attention to the parts of a policy most relevant to GDPR; this tool attempts to enhance compliance with GDPR's requirement for clear language compared to the legal language typically employed in privacy policies. Mousavi Nejad et al.~\cite{mousavi2018knight} developed a automated tool that uses deep learning to identify excerpts from privacy policies that map to specific GDPR articles. They found that on average their tool is 70-90\% accurate at identifying excerpts that are at least partially relevant to a GDPR article (significantly higher that a typical user); the authors argue that this tool is a significant step towards supporting user awareness of their rights under GDPR. Chang et al.~\cite{chang2019automated} developed a tool that employs machine learning to to extract the policy terms relevant to users' concerns and GDPR's requirements. In a user study with 96 participants, they found that their tool achieves .81 accuracy. 

Other tools include formal languages for expressing GDPR's transparency requirements~\cite{arfelt2019monitoring,grunewald2021tilt}, expressing and verifying privacy policies~\cite{tokas2022static}, and modeling inter-process communication to audit policies~\cite{basin2018purpose}. Wang et al. used cryptographic tools and trusted execution environments to automate compliance with privacy regulations~\cite{wang2022privguard}.

\subsubsection{Purpose or Processing Limitations}
%Discuss this, complicated.        
%28 papers
Ten laws impose purpose or processing limitations as a business obligation. These include both legal bases for processing and explicit purpose limitations. 

\vspace{5pt}
\paragraph{\it \hspace{-35pt} Legal bases for processing}
GDPR Article 6 enumerates six legal bases for processing, and this model has been replicated in many subsequent laws. One of those bases is user consent, turning a business obligation into a self managed right. Consent as a legal basis has been extensively studied by computer scientistis (Section~\ref{sec:consent}). 
Other legal bases have been studied less. Arfelt et al.~\cite{arfelt2019monitoring} express GDPR's legal basis requirement (Article 6(1)) in temporal logic and find it can be efficiently monitored. Kutylowski et al.~\cite{kutylowski2020gdpr} discuss challenges that can arise from GDPR's legal basis requirement in cases where a processor ceases to exist but personal data are still stored by a third-party storage provider.  Han et al. (2021)~\cite{han2021deeprec} argue that the ``Legitimate Interests'' clause in the GDPR covers use of data collected prior to GDPR; they build a system to provide sequential recommendations by training a global model using pre-GDPR data then fine-tuning that model locally % ie without sending the new data off the device
with more recent data. 

\vspace{5pt}
\paragraph{\it \hspace{-34pt} Purpose Limitation}
A few projects works have looked at purpose limitations in other contexts. 
One work found that after GDPR went into effect, Android apps declared fewer dangerous permissions  and that many apps reduced the number of times they accessed permissions~\cite{momen2019did}, suggesting that GDPR's purpose limitation requirement might have significantly enhanced privacy.  Another study found that additional metadata required to enforce purpose limitation and other GDPR requirements imposed a 2-5x performance slowdown on three widely-used database systems~\cite{shastri13understanding}. 

To facilitate compliance with purpose limitation requirements, Wolf et al. developed HivePBAC, an adaptation of purpose-based access control designed to ensure purpose limitation for message-oriented architectures~\cite{wolf2021messaging}. Karami et al. develop a programming language that generates runtime errors if data are used for purposes other than those for which they were collected or if they are not deleted after their purpose is complete~\cite{karami2022dpl}.

\subsubsection{Data Minimization}
%3.4.4
%9 papers
Ten laws %including GDPR, PIPA, PDPA, PIPEDA, LGPD and CPRA 
impose a data minimization obligation, %COPPA also requires companies to not condition participation on more data collection than necessary and to delete data that is no longer necessary. 
however this obligation has been rarely studied. One study found that personalization can be relatively robust to global minimization but that quality loss is significant for some users~\cite{biega2020operationalizing}.  
Anothers found that few apps enable  data-minimization SDK settings~\cite{kollnig2022goodbye} and that 16/31 IoT devices had at least one unnecessary data flow~\cite{mandalari2021blocking}. Senarath et al.~\cite{senarath2018understanding} found that developers struggle to implement data minimization due to uncertainty about how data could potentially be used, and that developers are inconsistent in how they applied data minimization. 

Arfelt et al.~\cite{arfelt2019monitoring} express a data minimization requirement in temporal logic and find it can be efficiently monitored. 
Shanmugam et al. proposed a framework for implementing data minimization in machine learning systems by iteratively estimating the system performance curve and use of personal data when a performance-based stopping criteria is achieved~\cite{shanmugam2022learning}. Zhou et al.~\cite{zhou2023policycomp} developed a tool for automatically comparing privacy policies between counterparts and identifying overly-broad data practices; their analysis flagged 48.3\% of policies as overly broad.

\subsubsection{Security Requirements}
%3.4.5        
%11 papers
Ten laws impose security requirements such as encryption for personal information. 
% GDPR encrypt
%GDPR requires personal information to be encrypted to ensure the security of personal data that is processed. %(Chapter 4 article 32). 
Longitudinal studies observed a 9\% decrease in plaintext transmission of personal data after GDPR went into effect, but of 39\% of top Android apps still transmitted plaintext data~\cite{jia2019leaks,fan2020empirical}. Many mHealth apps that attempted to encrypt data contained at least one error~\cite{fan2020empirical}. Although some apps contain geodifferences in security settings, those differences do not correspond to privacy regulation jurisdictions~\cite{kumar2022large}. 

In some cases, compliance with legally-mandated encryption requirements can impose a significant performance overhead~\cite{shah2019analyzing}. 
Marjanov et al. analyzed fines imposed for violations of GDPR's security requirements and identified common failings and danger points~\cite{marjanov2023data}.

\subsubsection{Privacy by Design}
%3.4.6        
%13 papers
Among the laws we reviewed, only GDPR includes a privacy by design obligation, although this obligation is also present in other laws beyond the scope of our review (e.g., Australia's Privacy Act, Kenya's Data Protection Act). A few projects have considered this requirement from various approaches. 
% developers 
Research has analyzed the mobile ecosystem through the lens of privacy by design~\cite{castelluccia2017privacy}, conducted case studies~\cite{gkotsopoulou2019data}, and studied developer perspectives on privacy by design to identify barriers~\cite{alkhatib2020privacy}. Recommendations include more guidance for developers~\cite{castelluccia2017privacy}, changes at individual and organization levels~\cite{alkhatib2020privacy}, and development of tools targeted at the engineering mindset~\cite{martin2018methods}. 

% tools
Independent work has developed tools for facilitating privacy by design using domain-specific languages~\cite{gerl2018lpl,gerl2019layered}, semantic models and automated verification~\cite{chhetri2022data}, and formal modeling and interactive theorem proving~\cite{kammueller2018formal}. Deshpande proposed an architecture for private-by-design database systems~\cite{deshpande2021sypse}. % using pseuodonymization and partitioning~\cite{deshpande2021sypse}. 
Tamò-Larrieux et al. analyze Privacy by Design as a stepping stone toward a right to customize data processing~\cite{tamo2021right}. 
        
\subsubsection{Record Keeping}
%3.4.7 
%2 papers
Nine laws impose a record-keeping requirement, but this obligation has been studied only in limited contexts such as measuring overhead incurred by adding synchronous logging to Redis~\cite{shah2019analyzing} or by a multi-level logging scheme~\cite{tran2021analyzing}. %Shah et al. modified Redis to comply with GDPR's record keeping requirement by adding synchronous logging and found that it incurred a  6-20x decrease in throughput (depending on tolerance for batching)~\cite{shah2019analyzing}.
%Tran et al. evaluated a multi-level logging scheme that complies with GDPR's record keeping requirements and found significant but manageable increases in CPU utilization (from 4\% to 8\%)~\cite{tran2021analyzing}. 
Ryan et al. proposed DPCat, a standardized representation for the collection and transfer of Register of Processing Activities (ROPA) information~\cite{ryan2022dpcat}. 
        
\subsubsection{Cross-Border Transfer Limitations}
%3.4.8
%5 papers
Nine laws impose restrictions on cross-border transfers. 
Analyses of the mobile tracking ecosystem prior to GDPR predicted a significant impact~\cite{razaghpanah2018apps}. However, subsequent work found that tracking flows did not change significantly~\cite{iordanou2018tracing} and that compliance rates were low, with  with 93\% of websites embedding third parties located in regions outside the Privacy Shield~\cite{urban2020beyond}  and 66\% of apps including cross-border transfers that were not accurately disclosed in their privacy policy~\cite{guaman2021gdpr}. IoT companion apps also transmit data across regions in ways that could violate GDPR~\cite{nan2023you}. 

%\subsubsection{Data Fiduciary Duties} % No laws
%empty
        
\subsubsection{Risk Assessments}
%3 papers
Eight laws require risk assessments such as Data Protection Impact Assessments, Privacy Impact Assessments, or Algorithmic Impact Assessments. Three projects have briefly considered this requirement in the context of big data systems~\cite{gruschka2018privacy}, database systems~\cite{shastri2019seven}, and smart homes~\cite{gkotsopoulou2019data}, but it hasn't been a significant focus.
        
\subsubsection{Contracting Requirements}
%1 paper
Five laws have contracting requirements, e.g., for service providers or third-parties who process personal information. Amaral et al.~\cite{amaral2023nlp} developed an NLP approach to automating compliance checking for GDPR data processing agreements. No other work has looked at this obligation. 
        
\subsubsection{Breach Notification Requirements}
%2 papers
Eight laws have breach notification requirements. Shastri et al.'s work on GDPR-compliant database systems briefly mentions this obligation~\cite{shastri13understanding,shastri2019seven}, but no work explicitly focuses on this requirement.

%%%%%%%%%%%
%%
%%  Applicability
%%
%%%%%%%%%%%

\subsection{Applicability}
The privacy laws and data protection regulations we analyzed included a range of different scopes of applicability, both in terms of which people were granted these protections (criteria included residence, citizenship, physical location, age, and employee status) and in terms of which organizations were subject to the regulation (criteria included number of users, company revenue, revenue from selling personal information, organization's country of registration, jurisdictions in which a company did business, and non-profit or governmental status). While some papers conducted cross-jurisdictional measurements or studies, none looked explicitly at the impact of legal applicability. 

%%%%%%%%%%%
%%
%%  General
%%
%%%%%%%%%%%

\subsection{General Papers}

While we were able to position most papers within our legal taxonomy, some work approached privacy regulations from a more general perspective. For example, a few projects explored generally what types of architectural and design changes would be required to bring systems into compliance with GDPR~\cite{gjermundrod2016privacytracker,pandit2018queryable,labadie2019understanding,hjerppe2019general,shastri2019seven,schwarzkopf2019position,bastos2018gdpr,campagna2020achieving}, and one did so for the proposed Indian Personal Data Protection Bill~\cite{singh2020technical}. Wong et al.~\cite{wong2023privacy} studied how GDPR and CCPA impact business risks identified in investor documents. Lopes et al.~\cite{lopes2019iso} evaluated how ISO 27001 standards might facilitate GDPR compliance, and 
Lu et al.~\cite{lu2021whois} considered GDPR's effect on WHOIS records. %They discuss the ICANN specifications on how to implement compliant redactions and properties of those specifications that limit or influence WHOIS providers in implementing these redactions. 
%Some papers looked at how various laws defined or interpreted terms. . 
Soussi et al.~\cite{soussi2020feasibility} explored the impact of GDPR on feasibility of large-scale vulnerability notifications. Truong et al.~\cite{truong2019gdpr} explored a blockchain-based approach to providing assurance of continuous GDPR-compliance. Others developed or proposed formal languages~\cite{barth2006privacy,arfelt2019monitoring,torre2019using,bonatti2020machine} or tools~\cite{elluri2018knowledge,ferrara2018static,piras2019defend,ferreira2023rulekeeper,aborujilah2022conceptual,bonatti2020machine,amaral2023nlp,tchana2023rgpdos,istvan2020towards} for facilitating and automating GDPR compliance. Other work explored users'~\cite{sheth2014us,zhang2021whether} and developers'~\cite{sirur2018we,freitas2018gdpr,chalhoub2020innovation,tahaei2021deciding,alomar2022developers,tahaei2022charting,chalhoub2022data,utz2023privacy,kekulluouglu2023we} attitudes towards, awareness of, and understanding of various privacy regulations. Since legal compliance is often framed as the developer's choice and responsibility~\cite{tahaei2021developers}, research has also explored advice or guidance available to (and used by) developers~\cite{tahaei2021developers,tahaei2022charting,castelluccia2017privacy,tahaei2022understanding,li2021developers,alomar2022developers} and identified barriers to compliance from the developer's perspective~\cite{tahaei2022charting,tahaei2021deciding,alomar2022developers,hadar2018privacy,bednar2019engineering,stover2023website,utz2023comparing,kekulluouglu2023we,tahaei2021privacy,alhazmi2020developers}. However, Utz et al.~\cite{utz2023comparing} found that notifications about privacy issues were less well-received than notifications about security issues.

Other work documented differences in various measures after a law went into effect. For example, Wang et al. found that 1/42 Android app libraries that exhibited cross-library data harvesting eliminated this behavior after GDPR went into effect~\cite{wang2021understanding}. 
Some work evaluated the impact of laws using cross-jurisdictional comparisons, for example finding now differences in universities' cloud migration between GDPR countries and others~\cite{fiebig2022heads}.  

\section{Discussion}\label{sec:discussion}

Looking at the 270 papers systematized in this work through an interdisciplinary lens, we see a deep, active, and highly impactful body of scholarship. However, systematizing this volume of work also illuminates patterns in how computer scientists currently approach the problem of analyzing the implementation and impact of privacy regulations. Based our analysis of these patterns and our systematization of the existing research in this space, we formulate recommendations about directions for future computer science research at the intersection of privacy and law. 

% get over self-management
\begin{rec}
Computer science researchers should recognize the growing inter-disciplinary consensus that privacy self-management is inherently flawed and move beyond critiques of self-management features.
\end{rec}

A majority of the work we systematized (51.5\%) focused on measuring, evaluating, or enhancing one or more privacy self-management features such as access, deletion, opting-in, or opting-out. Another 27.8\% of the papers focused on the right to transparency, and many of these were framed within the context of notice and consent, implicitly interpreting this requirement through the lens of privacy self-management; most of this work is comprised of criticisms of privacy self-management, such as quantifying designs that deter users from invoking their right or measuring the (poor) usability of existing implementations of self-management rights. 

The critiques of self-management resulting from the systematized work contribute to a growing interdisciplinary consensus that privacy self-management as not only inherently unworkable, but actually detrimental to privacy. Enhanced transparency is intended to empower users to make informed decisions about whether to consent to data practices, but the resulting disclosures are still unreadable~\cite{amos2021privacy,oh2021will} and omit critical information~\cite{fan2020empirical,yeratziotis2021complicy}. 
Implementations of self-management rights frequently deter people from invoking their rights by leveraging cognitive biases in so-called ``dark patterns''~\cite{draper2019corporate,waldman2020cognitive,gray2018dark,utz2019informed,o2021clear}. Moreover, self-management simply doesn't scale to the number of companies with which users regularly interact and the difficulty of identifying the many third parties with access to personal data~\cite{mcdonald2008cost,solove2012introduction,richards2018pathologies,solove2021myth}. 

Based on these compelling and consistent inter-disciplinary critiques, we believe computer scientists should consider critiques of self-management to be a solved problem. Instead of continuing generate new critiques of  self-management features, computer science researchers should evaluate non-self management features of existing regulations through measurements and user studies, and we should work to develop tools that complement regulatory efforts. 

% under-studied aspects
\begin{rec}
Computer science researchers should extend our efforts to evaluate the implementation and impact of currently under-studied aspects of privacy regulations. 
\end{rec}

Much of the work we systematized focused on a small number of legal aspects such as right to access (34 papers), consent interfaces (41 papers), and right to transparency (75 papers). By contrast, other  aspects of modern privacy laws have gone largely unstudied by our community. Some of these understudied aspects challenge and enhance common approaches to self-management or take users out of the self-management loop. Examples of under-studied legal aspects that could benefit from additional research include data minimization (9 current papers), anti-discrimination requirements (0 current papers), prohibitions on certain technologies such as facial recognition (1 current paper), and rights not to be subject to automated decisions (4 current papers). 

\begin{rec}
Computer science researchers should prioritize longitudinal and cross-cultural work at the intersection of technology and privacy law. 
\end{rec}

The measurement work systematized in this paper focused on the time immediately after or immediately around the date when a law went into effect. However, the legal community considers post-enactment guidelines and case law to provide essential interpretations of laws that define the rights and obligations imposed by those laws. For example, what constitutes a valid opt-out of sale mechanism under California law has evolved since January 2020 via new guidelines, new interpretations and legal actions, and subsequent legal amendments. To provide meaningful evaluations of privacy regulations, future work will need to conduct longitudinal measurement studies over longer periods after a law goes into effect, specifically including the evaluation of post-enactment events that can change legal interpretation. 

Similar sounding legal rights can also result substantively legal realities in different jurisdictions due to different interpretations in different jurisdictions. For example, many national privacy laws provide special protection for ``sensitive personal data'', but Singapore's PDPA provides no statutory definition or protection for sensitive data over other types of data~\cite{pdpa}. However, guidance from Singapore's Personal Data Protection Commission indicates that sensitivity of data is a factor for consideration, thus introducing the potential for sensitive data protections under a law which otherwise lacks such protections~\cite{pdpc2020advisory}. These differences can result in results that don't generalize between different jurisdictions even if they have similar legal aspects and regulatory language. For example, consider 
an illustrative example from the research on dark patterns and nudging: two studies investigated the effects of the on-page location of consent banners on interactions and  user consent  decisions~\cite{utz2019informed,bermejo2021website}. A study with Europeans found  participants were three times more likely to interact with banners in the lower-left than in other positions; a study with a predominantly-American population found  position had no effect on interaction~\cite{bermejo2021website}. While there are many possible reasons for these inconsistencies (including timing, recruitment, and methodology), one possibility is that a genuine  difference exists between these two populations, which may have been induced by cultural factors, differing experiences with privacy and consent, or the impact that the GDPR (and the changes to corporate behavior it triggered, e.g., the frequency of cookie banners) might have had on European's behavior. To provide meaningful results about the impact of privacy regulations globally, researchers should conduct cross-jurisdictional studies and should replicate results from single-jurisdictional work to validate whether the results generalize to other legal jurisdictions. 

% Interdisciplinary
%\begin{rec}
%Computer science researchers should work with interdisciplinary teams that bring diverse perspectives to problems relating to Internet privacy and data protection. 
%\end{rec}

% Future regulations
\begin{rec}
Computer science researchers should explore how technical expertise and methodology might be applied to evaluating proposed future regulations and regulatory approaches in addition to laws currently in effect. \end{rec}

Existing work provides critiques of extant laws. This can (and has) impacted subsequent interpretation and enforcement. However, this impact  could be amplified if computer scientists developed  techniques to empirically evaluate proposed regulations. Certain upcoming and proposed laws contain absolute clauses beyond the scope of existing privacy guarantees, for example the District of Columbia's Stop Discrimination by Algorithms Act of 2021 and Section 207 ("Civil Rights and Algorithms") of the proposed new US omnibus privacy legislation, the American Data Privacy and Protection Act (ADPPA). Such laws are beginning to address a key facet of privacy regulation, which is the ways in which it can be made to enforce equity in privacy across axes of marginalization and class. 
Studies of  these laws would be valuable for understanding how non-discrimination approaches impact privacy outcomes and  privacy equity. 

Finally, we encourage the study of regulatory regimes which have only been theorized in academic law literature, such as data fiduciary approaches~\cite{balkin2020fiduciary} and civil rights approaches~\cite{bedoya2020privacy}. Such regimes could  be deeply transformative but would rely on major shifts in both privacy regulations and also how societies construct and value privacy. 
\newpage

%\begin{acks}
%To Robert, for the bagels and explaining CMYK and color spaces.
%\end{acks}

\bibliographystyle{plain}
\bibliography{main}

%\appendix
%\input{sections/90-appendix}

\end{document}